\documentclass[a4paper,11pt]{article}
\pdfoutput=1 

\usepackage{template} 

\usepackage{amssymb}
\usepackage{amsmath}
\usepackage{bbold}
\usepackage{slashed}
\usepackage{graphicx}
\usepackage[compat=1.1.0]{tikz-feynman}
\usepackage{cancel}
\usepackage{tensor}
\usepackage{epstopdf}
\usepackage{tikz}
\usepackage{colortbl,hhline}
\usepackage{hyperref}
\usepackage{subcaption}

\newcommand{\ol}{\overline}

\title{\boldmath Neutrino mixing sum rules and the Littlest Seesaw}

\author[a]{Francesco Costa}
\author[b]{Stephen F. King}

\affiliation[a]{Institute for Theoretical Physics, 
	Georg-August University G\"ottingen,\\
	Friedrich-Hund-Platz 1, G\"ottingen, D-37077 Germany}
\affiliation[b]{School of Physics and Astronomy, University of Southampton,\\
	Southampton SO17 1BJ, United Kingdom}

\emailAdd{francesco.costa@theorie.physik.uni-goettingen.de}
\emailAdd{ s.f.king@soton.ac.uk }

\abstract{In this work, we study the neutrino mixing sum rules arising from discrete symmetries, and the class of Littlest Seesaw (LS) neutrino models. These symmetry based approaches 
all offer predictions for the cosine of the leptonic CP phase $\cos \delta$ in terms of the mixing angles,
$\theta_{13}$, $\theta_{12}$, $\theta_{23}$, while the LS models also predict the sine of the leptonic CP phase $\sin \delta$
as well as making other predictions. In particular we study the \textit{solar} neutrino mixing sum rules, arising from charged lepton corrections to Tri-bimaximal (TB), Bi-maximal (BM), Golden Ratios (GRs) and Hexagonal (HEX) neutrino mixing,
and \textit{atmospheric} neutrino mixing sum rules, arising from preserving one of the columns of these types of mixing, for example the first or second column of the TB mixing matrix (TM1 or TM2),
and confront them with an up-to-date global fit of the neutrino oscillation data.  We show that some mixing sum rules, for example an \textit{atmospheric} neutrino mixing sum rule
arising from a version of neutrino Golden Ratio mixing (GRa1), are already excluded at 3$\sigma$, and determine the remaining models allowed by the data. We also consider the more predictive LS models (which obey the TM1 sum rules and offer further predictions) based on constrained sequential dominance
CSD($n$) with $n\approx 3$. We compare for the first time the three cases
$n=2.5$, $n=3$ and $n=1+\sqrt{6}\approx 3.45$ which are favoured by theoretical models,
using a new type of analysis to accurately predict the observables $\theta_{12}$, $\theta_{23}$ and $\delta$.
We study all the above approaches, \textit{solar} and \textit{atmospheric} mixing sum rules and LS models,
together so that they may be compared, and to give an up to date analysis of the predictions of all of these possibilities, when confronted with the most recent global fits. 
}

\begin{document} 
\maketitle
\flushbottom

\section{Introduction}

Neutrino mass and mixing represents the first and so far the only new physics beyond the Standard Model (SM) of particle physics. We know it must be new physics because its origin is unknown and it is not predicted by the SM.
Independently of the whatever the new (or nu) SM is, we do know that the minimal paradigm involves three active neutrinos, the weak eigenstates $\nu_e, \nu_{\mu}, \nu_{\tau}$ (the $SU(2)_L$ partners to the left-handed charged lepton mass eigenstates) which are related to the three mass eigenstates $m_{1,2,3}$ by a unitary PMNS mixing matrix \cite{Workman:2022ynf}.

The PMNS matrix is similar to the CKM matrix which describes quark mixing, but involves three independent leptonic mixing angles $\theta_{23}, \theta_{13}, \theta_{12}$
(or $s_{23}=\sin \theta_{23}$, $s_{13}=\sin \theta_{13}$, $s_{12}=\sin \theta_{12}$),
one leptonic CP violating Dirac phase $\delta$ which affects neutrino oscillations, and possibly two Majorana phases which do not enter into neutrino oscillation formulas. 
Furthermore neutrino oscillations only depend on the two mass squared differences $\Delta m^2_{21}=m_2^2-m_1^2$, which is constrained by data to be positive, and $\Delta m^2_{31}=m_3^2-m_1^2$, which current data allows to take a positive (normal) or negative (inverted) value. 
In 1998, the angle $\theta_{23}$ was first measured to be roughly $45^{\circ}$ \cite{Super-Kamiokande:1998kpq} (consistent with equal bi-maximal 
$\nu_{\mu}- \nu_{\tau}$ mixing)
by atmospheric neutrino oscillations, while $\theta_{12}$ was determined to be roughly $35^{\circ}$ 
(consistent with equal tri-maximal $\nu_e - \nu_{\mu}- \nu_{\tau}$ mixing)
in 2002 by solar neutrino oscillation experiments \cite{SNO:2001kpb}, while 
$ \theta_{13}$ was first accurately found to be $8.5^{\circ}$ in 2012 by reactor oscillation experiments \cite{DayaBay:2012fng,RENO:2012mkc}.

Various simple ansatzes for the PMNS matrix were proposed, the most simple ones involving a zero reactor angle 
and bimaximal atmospheric mixing, $s_{13}=0$ and $s_{23}=c_{23}=1/\sqrt{2}$,
leading to a PMNS matrix of the form,
\begin{eqnarray}
U_0 =
\left( \begin{array}{ccc}
c_{12} & s_{12} & 0\\
 -\frac{s_{12}}{\sqrt{2}}  &  \frac{c_{12}}{\sqrt{2}} & \frac{1}{\sqrt{2}} \\
\frac{s_{12}}{\sqrt{2}}  &  -\frac{c_{12}}{\sqrt{2}} & \frac{1}{\sqrt{2}} 
\end{array}
\right),
\label{GR}
\end{eqnarray}
where the zero subscript reminds us that this form has $\theta_{13}=0$ (and $\theta_{23}=45^\circ$).

For golden ratio (GRa) mixing~\cite{Datta:2003qg}, 
the solar angle is given by
$\tan \theta_{12}=1/\phi$, where $\phi = (1+\sqrt{5})/2$ is the golden ratio
which implies $\theta_{12}=31.7^\circ$. 
There are two alternative versions where 
$\cos \theta_{12} =\phi/2$ and $\theta_{12}=36^\circ$~\cite{Rodejohann:2008ir}
which we refer to as GRb mixing, and GRc where $\cos \theta_{12} =\phi/\sqrt{3}$ and $\theta_{12} \approx 20.9^{\circ}$.

For bimaximal (BM) mixing (see e.g.~\cite{Davidson:1998bi,Altarelli:2009gn,Meloni:2011fx} and references therein), 
we insert $s_{12}=c_{12}=1/\sqrt{2}$ ($\theta_{12}=45^\circ$) into Eq.~(\ref{GR}), 
\begin{eqnarray}
U_{\mathrm{BM}} =
\left( \begin{array}{ccc}
\frac{1}{\sqrt{2}} & \frac{1}{\sqrt{2}} & 0\\
-\frac{1}{2}  & \frac{1}{2} & \frac{1}{\sqrt{2}} \\
\frac{1}{2}  & -\frac{1}{2} & \frac{1}{\sqrt{2}} 
\end{array}
\right).
\label{BM}
\end{eqnarray}

For tri-bimaximal (TB) mixing~\cite{Harrison:2002er}, alternatively we use
$s_{12}=1/\sqrt{3}$, $c_{12}=\sqrt{2/3}$ ($\theta_{12}=35.26^\circ$) in Eq.~(\ref{GR}),
\begin{eqnarray}
U_{\mathrm{TB}} =
\left( \begin{array}{ccc}
\sqrt{\frac{2}{3}} & \frac{1}{\sqrt{3}} & 0 \\
-\frac{1}{\sqrt{6}}  & \frac{1}{\sqrt{3}} & \frac{1}{\sqrt{2}} \\
\frac{1}{\sqrt{6}}  & -\frac{1}{\sqrt{3}} & \frac{1}{\sqrt{2}}
\end{array}
\right).
\label{TB}
\end{eqnarray}
Finally another pattern studied in the literature with $\theta_{13}=0$ (and $\theta_{23}=45^\circ$) is the hexagonal mixing (HEX) where $\theta_{12} = \pi/6$. 

These proposals are typically enforced by finite discrete symmetries such as $A_4, S_4, S_5$ (for a review see e.g. 
\cite{King:2013eh}). After the reactor angle was measured, which excluded all these ansatze, 
there were various proposals to rescue them and to maintain the notion of predictivity of the leptonic mixing parameters. 
Indeed the measurement of the reactor angle opens up the possibility to predict the CP phase $\delta$, which is not directly measured so far and remains poorly determined even indirectly.
Two approaches have been developed, in which some finite symmetry (typically a subgroup of $A_4, S_4, S_5$) can enforce a particular structure of the PMNS matrix consistent with a non-zero reactor angle, leading to \textit{solar} and \textit{atmospheric} sum rules, as we now discuss.

The first approach, which leads to \textit{solar} sum rules, is to assume that the above patterns of mixing still apply to the neutrino sector, but receive charged lepton mixing corrections due to the PMNS matrix being the product of two unitary matrices, which in our convention is written as  
$V_{eL} V^{\dagger}_{\nu_L}$, where $V^{\dagger}_{\nu_L}$ is assumed to take the BM, TB or GR form, while $V_{eL}$ differs from the unit matrix.
If $V_{eL}$ involves negligible 13 charged lepton mixing, then it is possible to generate a non-zero 13 PMNS mixing angle, while leading to correlations amongst the physical PMNS parameters, known as {\em solar} mixing sum rules
\cite{King:2005bj,Masina:2005hf,Antusch:2005kw,Antusch:2007rk}.
This scenario may be enforced by a subgroup of $A_4, S_4, S_5$ which enforces the $V_{\nu}$ structure \cite{King:2013eh} while allowing charged lepton corrections.

In the second approach, which leads to \textit{atmospheric} sum rules,
it is assumed that the physical PMNS mixing matrix takes the BM, TB or GR form but only in its first or second column, while the third column necessarily departs from these structures due to the non-zero 13 angle.
Such patterns again lead to correlations amongst the physical PMNS parameters, known as 
{\em atmospheric} mixing sum rules. This scenario may be enforced by a subgroup of $A_4, S_4, S_5$ which enforces the one column $V_{\nu}$ structure \cite{King:2013eh} while forbidding charged lepton corrections.

Apart from the large lepton mixing angles, another puzzle is the extreme lightness of neutrino masses.
Although the type I seesaw mechanism can qualitatively explain the smallness of neutrino masses through the heavy right-handed neutrinos (RHNs), if one doesn't make other assumptions, it contains too many parameters to make any particular predictions for neutrino mass and mixing. The sequential dominance (SD)~\cite{King:1998jw,King:1999cm} of right-handed neutrinos proposes that 
the mass spectrum of heavy Majorana neutrinos is strongly hierarchical, i.e. $M_\text{atm}\ll M_\text{sol}\ll M_\text{dec}$,
where the lightest RHN with mass $M_\text{atm}$ is 
responsible for the atmospheric neutrino mass, that with mass $M_\text{sol}$ gives 
the solar neutrino mass, and a third largely decoupled RHN
gives a suppressed lightest neutrino mass. It leads to an effective two right-handed neutrino (2RHN) model ~\cite{King:1999mb,Frampton:2002qc} with a natural explanation for the physical neutrino mass hierarchy, with normal ordering and the lightest neutrino being approximately massless, $m_1=0$.

A very predictive minimal seesaw model with two right-handed neutrinos and one texture zero is the so-called constrained sequential dominance (CSD) model~\cite{King:2005bj,Antusch:2011ic,King:2013iva,King:2015dvf,King:2016yvg,Ballett:2016yod,King:2018fqh,King:2013xba,King:2013hoa,Bjorkeroth:2014vha}.
The CSD($n$) scheme, also known as the Littlest Seesaw,
assumes that the two columns of the Dirac neutrino mass matrix are proportional to $(0,1, -1)$ and $(1, n, 2-n)$ or
$(1, 2-n, n)$ respectively in the RHN diagonal basis (or equivalently $(0,1, 1)$ and $(1, n, n-2)$ or $(1, n-2, n)$)
where the parameter $n$ was initially assumed to be a positive integer, but in general may be a real number. For example the CSD($3$) (also called Littlest Seesaw model)~\cite{King:2013iva,King:2015dvf,King:2016yvg,Ballett:2016yod,King:2018fqh}, CSD($4$) models~\cite{King:2013xba,King:2013hoa} and CSD($2.5$)~\cite{Chen:2019oey} can give rise to phenomenologically viable predictions for lepton mixing parameters and the two neutrino
mass squared differences $\Delta m^2_{21}$ and $\Delta m^2_{31}$,
corresponding to special constrained cases of lepton mixing which preserve the first column of the TB mixing matrix,
namely TM1 and hence satisfy \textit{atmospheric} mixing sum rules. 
As was observed, modular symmetry remarkably suggests CSD($1+\sqrt{6}$) $\approx$ CSD($3.45$)~\cite{Ding:2019gof,Ding:2021zbg,deMedeirosVarzielas:2022fbw,deAnda:2023udh}.

In this paper we study neutrino \textit{solar} and \textit{atmospheric} mixing sum rules arising from discrete symmetries, and also discuss the class of Littlest Seesaw (LS) models corresponding to CSD($n$) with $n\approx 3$.
The motivation is to study all the above symmetry based approaches, 
namely \textit{solar} and \textit{atmospheric} mixing sum rules and LS models, together in one place so that they may be compared, and to give an up to date analysis of the predictions of all of these possibilities, when confronted with the most recent global fits. 
All these approaches offer predictions for the cosine of the leptonic CP phase $\cos \delta$ in terms of the mixing angles,
$\theta_{13}$, $\theta_{12}$, $\theta_{23}$,
which can be tested in forthcoming high precision neutrino experiments. 
In particular we study the \textit{solar} neutrino mixing sum rules, arising from charged lepton corrections to TB, BM and GR neutrino mixing,
and \textit{atmospheric} neutrino mixing sum rules, arising from preserving one of the columns of these types of mixing, for example the first or second column of the TB mixing matrix (TM1 or TM2),
and confront them with an up-to-date global fit of the neutrino oscillation data.  We show that some mixing sum rules, for example all the \textit{atmospheric} neutrino mixing sum rule
arising from a Golden Ratio mixings are already excluded at 3$\sigma$ a part from GRa2, and determine the remaining models allowed by the data. We also give detailed comparative results for the highly predictive LS models 
(which are special cases of TM1).
These models are highly predictive with only two free real parameters fixing all the neutrino oscillation observables, making them candidates for being the most minimal predictive seesaw models of leptons still compatible with data. This is the first time that the three LS cases corresponding to CSD($n$) with
$n=2.5$, $n=3$ and $n=1+\sqrt{6}\approx 3.45$ have been studied together in one place, using the most up to date global fits. These three cases are predicted by theoretical models. In particular $n=3$ was studied in a flavon model based on $S_4$ ~\cite{King:2013iva,King:2015dvf,King:2016yvg,Ballett:2016yod,King:2018fqh}, $n=2.5$ was introduced in the tri-direct CP approach based on the flavour symmetry $S_4 \times Z_5 \times Z_8$~\cite{Chen:2019oey}, and $n=1+\sqrt{6}\approx 3.45$ derived in the modular symmetry framework with three $S_4$ groups~\cite{Ding:2019gof,Ding:2021zbg,deMedeirosVarzielas:2022fbw,deAnda:2023udh}.
We also propose a new way of analysing these models, which allows accurate predictions for the least well determined oscillation parameters $\theta_{12}$, $\theta_{23}$ and $\delta$ to be extracted.

The layout of the remainder of the paper is as follows. In Chapter \ref{chap:2} we introduce the notation for the PMNS matrix and discuss the symmetries of the leptonic Lagrangian. In Chapter \ref{chap:3} and \ref{chap:4} we introduce the \textit{atmospheric} and \textit{solar} sum rules for the different models we are studying and confront them with the up-to-date neutrino data global fit. We proceed in Chapter \ref{chap:5} discussing the CDS and the Littlest Seesaw model, showing its high predictivity and the viable parameter space given the experimental data and its fit. Finally we conclude in Chapter \ref{chap:6}.


\section{Lepton mixing and symmetries}
\label{chap:2}

The mixing matrix in the lepton sector, the PMNS matrix
$U_{\mathrm{PMNS}}$, is defined as the matrix which appears in the
electroweak coupling to the $W$ bosons expressed in terms of lepton
mass eigenstates. With the mass matrices of charged leptons
$M_{e}$ and neutrinos $M^{\nu}_{LL}$ written as\footnote{Although we
have chosen to write a Majorana mass matrix, all relations in the
following are independent of the Dirac or Majorana nature of neutrino
masses.}
\begin{eqnarray}
L=-  \ol{e_L} M^{e} e_R  
- \frac{1}{2}\ol{\nu_L} M^{\nu} \nu_{L}^c 
+ {H.c.}\; ,
\end{eqnarray}
and performing the transformation from flavour to mass basis by
 \begin{eqnarray}\label{eq:DiagMe}
V_{e_L} \, M^{e} \,
V^\dagger_{e_R} =
\mbox{diag}(m_e,m_\mu,m_\tau)
 , \quad~
V_{\nu_L} \,M^{\nu}\,V^T_{\nu_L} =
\mbox{diag}(m_1,m_2,m_3),
\label{mLLnu}
\end{eqnarray}
the PMNS matrix is given by
\begin{eqnarray}\label{Eq:PMNS_Definition}
U_{\mathrm{PMNS}}
= V_{e_{L}} V^\dagger_{\nu_{L}} \,.
\end{eqnarray}
Here it is assumed implicitly that unphysical phases are removed by
field redefinitions, and $U_\mathrm{PMNS}$ contains one Dirac phase
and two Majorana phases. The latter are physical only in the case of
Majorana neutrinos, for Dirac neutrinos the two Majorana phases can be
absorbed as well.

According to the above discussion, the neutrino mass and flavour bases are misaligned by the PMNS matrix
as follows,

\begin{align}
	\begin{aligned}
		\left(\begin{array}{c}
			\nu_e  \\
			\nu_{\mu} \\
			\nu_{\tau} 
		\end{array}\right) = & \left(\begin{array}{ccc}
		U_{e1} & U_{e2} & U_{e3} \\
		U_{\mu 1} & U_{\mu 2} & U_{\mu 3} \\
		U_{\tau 1} & U_{\tau 2} & U_{\tau 3} 
		\end{array}\right) 
		 \left(\begin{array}{c}
			\nu_1  \\
			\nu_{2} \\
			\nu_{3} 
		\end{array}\right) \equiv U_{\text {PMNS }} \left(\begin{array}{c}
		\nu_1  \\
		\nu_{2} \\
		\nu_{3} 
		\end{array}\right),
	\end{aligned}
\end{align}
where $\nu_e, \nu_{\mu}, \nu_{\tau}$ are the $SU(2)_L$ partners to the left-handed charged lepton mass eigenstates and
$\nu_{1,2,3}$ are the neutrinos in their mass basis. Following the standard convention we can describe $U_{\text {PMNS }}$ in terms of three angles, one CP violation phase and two Majorana phases

\begin{align}
	\begin{aligned}
		U_{\text {PMNS }}= & \left(\begin{array}{ccc}
			1 & 0 & 0 \\
			0 & c_{23} & s_{23} \\
			0 & -s_{23} & c_{23}
		\end{array}\right)\left(\begin{array}{ccc}
			c_{13} & 0 & s_{13} e^{-\mathrm{i} \delta} \\
			0 & 1 & 0 \\
			-s_{13} e^{\mathrm{i} \delta} & 0 & c_{13}
		\end{array}\right)  \left(\begin{array}{ccc}
		c_{12} & s_{12} & 0 \\
		-s_{12} & c_{12} & 0 \\
		0 & 0 & 1
		\end{array}\right) P,
	\end{aligned}
\end{align}

\begin{align}
	= \left(\begin{array}{ccc}
		c_{12} c_{13} & s_{12} c_{13} & s_{13} e^{-i \delta} \\
		-s_{12} c_{23}-c_{12} s_{13} s_{23} e^{i \delta} & c_{12} c_{23}-s_{12} s_{13} s_{23} e^{i \delta} & c_{13} s_{23} \\
		s_{12} s_{23}-c_{12} s_{13} c_{23} e^{i \delta} & -c_{12} s_{23}-s_{12} s_{13} c_{23} e^{i \delta} & c_{13} c_{23}
	\end{array}\right) P,
	\label{eq:parametrisation}
\end{align}
where $P$ contains the Majorana phases

\begin{align}
   P= 	 \operatorname{diag}\left(1, e^{i \alpha_{21} / 2}, e^{i \alpha_{31} / 2}\right),
\end{align}
The current $3 \sigma$ parameters intervals coming from the global fit of the neutrino oscillation data by the \hyperlink{http://www.nu-fit.org/}{nuFIT} collaboration \cite{Esteban:2020cvm} are

\begin{align}
&	\theta_{12}=[31.31^{\circ},35.74^{\circ}], \quad   \theta_{23}=[39.6^{\circ},51.9^{\circ}], \quad  \theta_{13}=[8.19^{\circ},8.89^{\circ}], \quad  \\& \delta = [0^{\circ},44^{\circ}]\quad \& \quad[108^{\circ},360^{\circ}], \quad  \frac{\Delta_{21}^2}{10^{-5} \mathrm{eV}^2} = [6.82,8.03], \quad   \frac{\Delta_{3l}^2}{10^{-3} \mathrm{eV}^2} = [2.428,2.597].
	\label{eq:exp-data}\quad \quad 
\end{align}
The PMNS matrix reads

\begin{align}
	|U|_{3 \sigma}^{\text {w/o SK-atm }}=\left(\begin{array}{lll}
		0.803 \rightarrow 0.845 & 0.514 \rightarrow 0.578 & 0.142 \rightarrow 0.155 \\
		0.233 \rightarrow 0.505 & 0.460 \rightarrow 0.693 & 0.630 \rightarrow 0.779 \\
		0.262 \rightarrow 0.525 & 0.473 \rightarrow 0.702 & 0.610 \rightarrow 0.762
	\end{array}\right).
\end{align}
These results are obtained considering normal ordering, which is the current best fit, and without including the Super-Kamiokande (SK) data.%
\begin{figure}[t]
	\centering
	\includegraphics[scale=0.8]{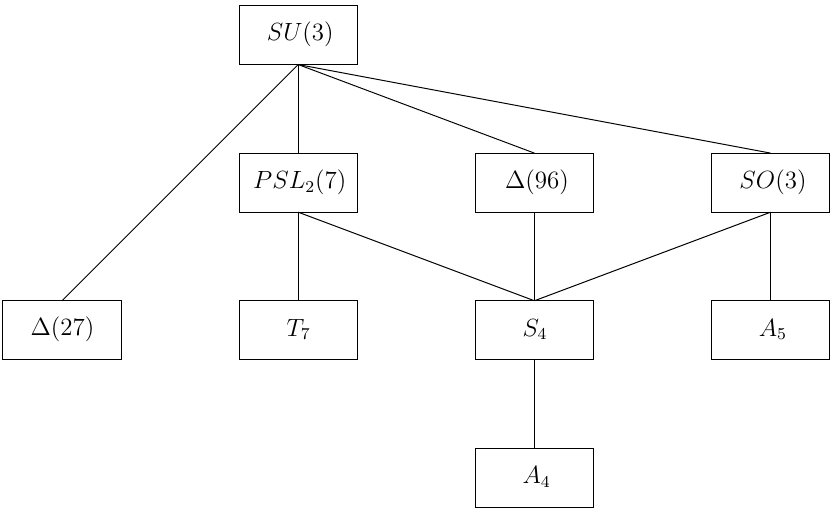}
	\caption{\small Subgroups of $SU(3)$ with triplet representations. The smaller of two groups connected in the graph is a subset of the other. Figure from \cite{King:2013eh}.}
	\label{fig:families}
\end{figure}
Simple mixing patter such TB, BM or GR could explain the first neutrino oscillation data. These patterns can be enforced via symmetries of the mass matrices. Let us take a basis where the charged lepton $M_e$ mass matrix is diagonal and we notice that for 3 generations we have that $Z^T_3$ is a symmetry of the Lagrangian

\begin{align}
	T^{\dagger}\left(M_e^{\dagger} M_e\right) T=M_e^{\dagger} M_e,
\end{align}
where $T=\operatorname{diag}\left(1, \omega^2, \omega\right)$ and $\omega=e^{i 2 \pi / 3}$. The light Majorana neutrino mass matrix is invariant under the Klein symmetry: $Z_2^U \times Z_2^S$. This can be seen taking the diagonal neutrino mass matrix and performing the transformations

\begin{align}
	M^\nu=S^T M^\nu S, \quad M^\nu=U^T M^\nu U,
\end{align}
and $M^\nu$ is left invariant with

\begin{align}
	\begin{gathered}
		S=U_{\mathrm{PMNS}}^* \operatorname{diag}(+1,-1,-1) U_{\mathrm{PMNS}}^T \\
		U=U_{\mathrm{PMNS}}^* \operatorname{diag}(-1,+1,-1) U_{\mathrm{PMNS}}^T,
	\end{gathered}
	\label{eq:SandU}
\end{align}
where this result follows from the fact that, in the charged lepton mass eigenstate basis, the neutrino mass matrix is diagonalised by $U_{\mathrm{PMNS}}$ as in Eq. \eqref{mLLnu}, where any two diagonal matrices commute. Then Eq. \eqref{eq:SandU} shows that the matrices $S,U$ are both diagonalised by the same matrix 
$U_{\mathrm{PMNS}}$ that also diagonalises the neutrino mass matrix. Given this result, we can always find 
the two matrices $S,U$ for any PMNS mixing matrix, and hence the Klein symmetry 
is present for any choice of the PMNS mixing. However not all Klein symmetries may be identified with
finite groups of low order.

This description is meaningful if the charged leptons are diagonal (T is conserved) or approximately diagonal (T is softly broken). We are therefore interested in finite groups that are superset of $Z_2^U \times Z_2^S$ and $Z_3^T$ and have a triplet representation. Groups of low order that satisfy these constraints are given in Figure \ref{fig:families}. 
\begin{figure}
	\centering
	\includegraphics[scale=0.15]{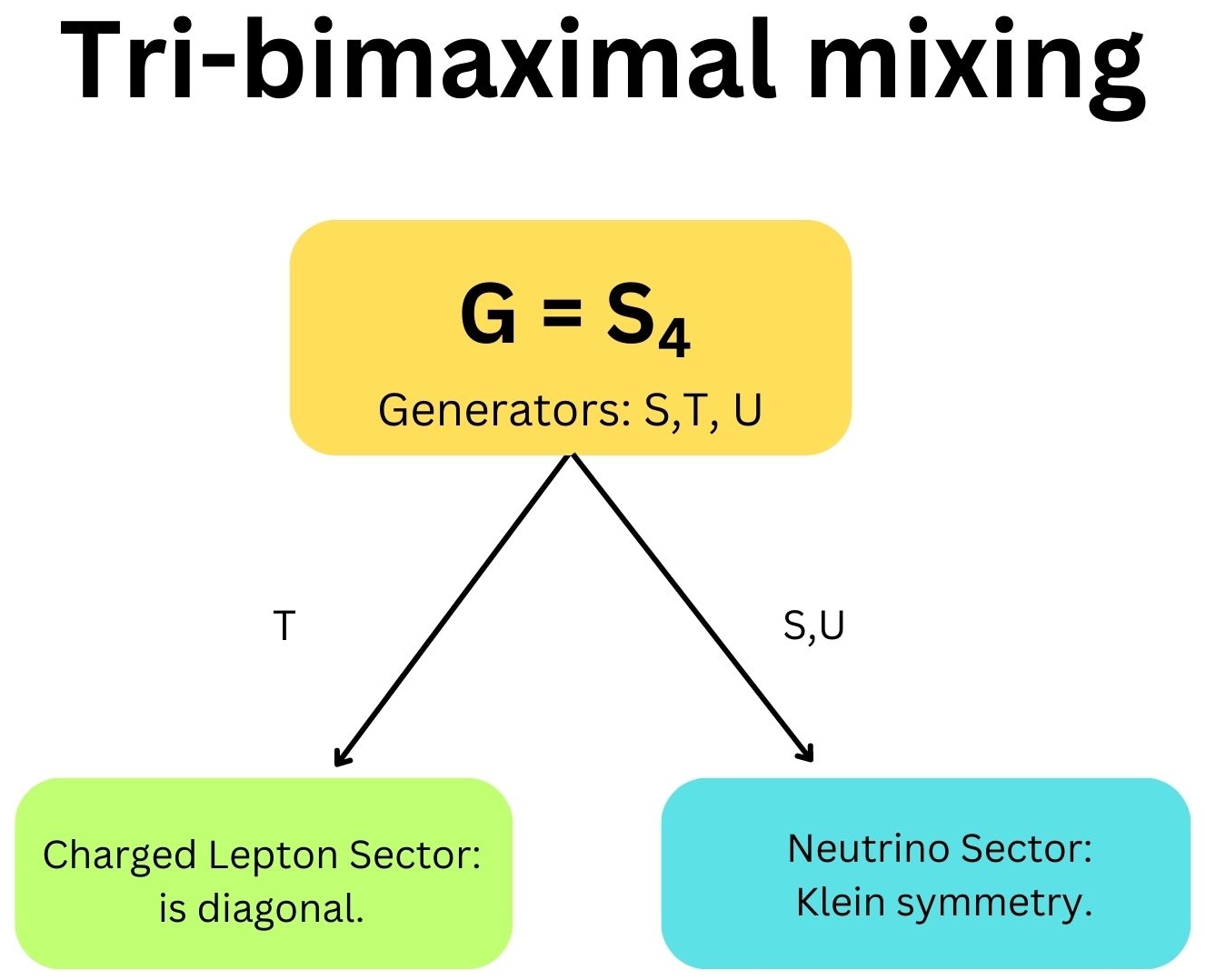}
	\caption{\small A schematic diagram that illustrate the way that the two subgroups $Z_2^U \times Z_2^S$ and $Z_3^T$ of a finite group work in the charged lepton and neutrino sectors in order to enforce a particular pattern of PMNS mixing. In this example, the group $S_4$ leads to TB mixing.}
	\label{fig:tb}
\end{figure}

One simple example is the group $G = S_4$, of order 24, which is the group of permutation of 4 objects. The generators follow the presentation rules \cite{King:2013eh}

\begin{align}
	S^2=T^3=(S T)^3=U^2=(T U)^2=(S U)^2=(S T U)^4=\mathbf{1},
	\label{eq:presentation}
\end{align}
The two possible $S_4$ triplet irreducible representations with a standard choice of basis \cite{King:2009ap}, gives the generators explicit expression 
 
\begin{align}
	S=\frac{1}{3}\left(\begin{array}{ccc}
		-1 & 2 & 2 \\
		2 & -1 & 2 \\
		2 & 2 & -1
	\end{array}\right), \quad T=\left(\begin{array}{ccc}
		1 & 0 & 0 \\
		0 & \omega^2 & 0 \\
		0 & 0 & \omega
	\end{array}\right), \quad U=\mp\left(\begin{array}{ccc}
		1 & 0 & 0 \\
		0 & 0 & 1 \\
		0 & 1 & 0
	\end{array}\right),
	\label{eq:generators-S4}
\end{align}
where again $\omega=e^{i 2 \pi / 3}$ and the sign of the $U$ matrix corresponds to the two different triplet representation. The group $S_4$ predicts a TB mixing \cite{Harrison:2002er}, see Figure \ref{fig:tb}. This can be checked by the fact that $S$ and $U$ are diagonalised by $U_{\mathrm{TB}}$, see Eqs. \eqref{eq:SandU}.
Another commonly used group is $A_4$, which has two generators $S$ and $U$ that follow the same presentation rules as in Eq. \eqref{eq:presentation} and in a standard basis \cite{Altarelli:2005yx}, the generators have the same form as in Eq. \eqref{eq:generators-S4}.

\begin{figure}[t]
	\centering
	\begin{subfigure}{.5\textwidth}
		\centering
		\includegraphics[scale=0.15]{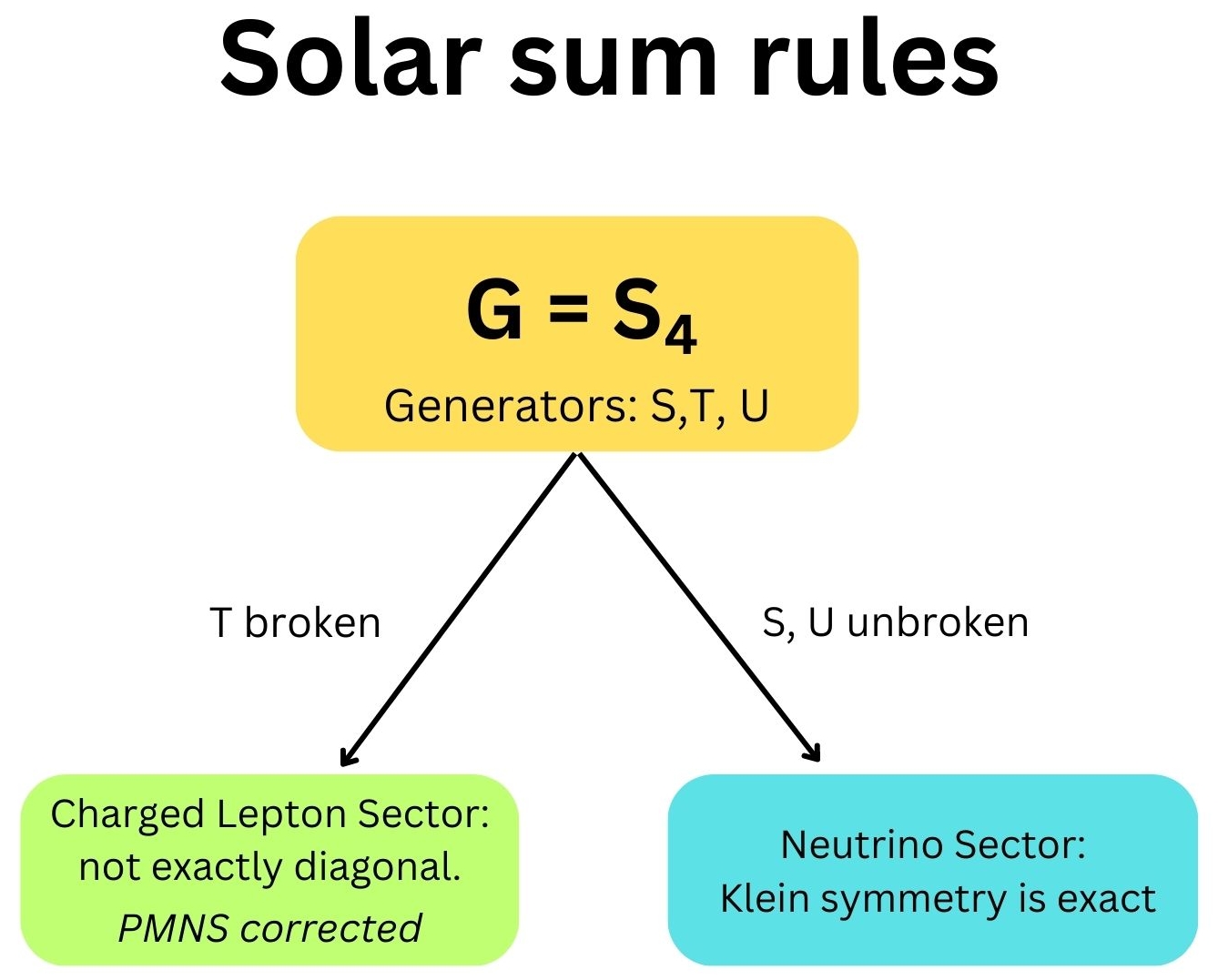}
		\label{fig:sub1}
	\end{subfigure}%
	\begin{subfigure}{.5\textwidth}
		\centering
		\includegraphics[scale=0.15]{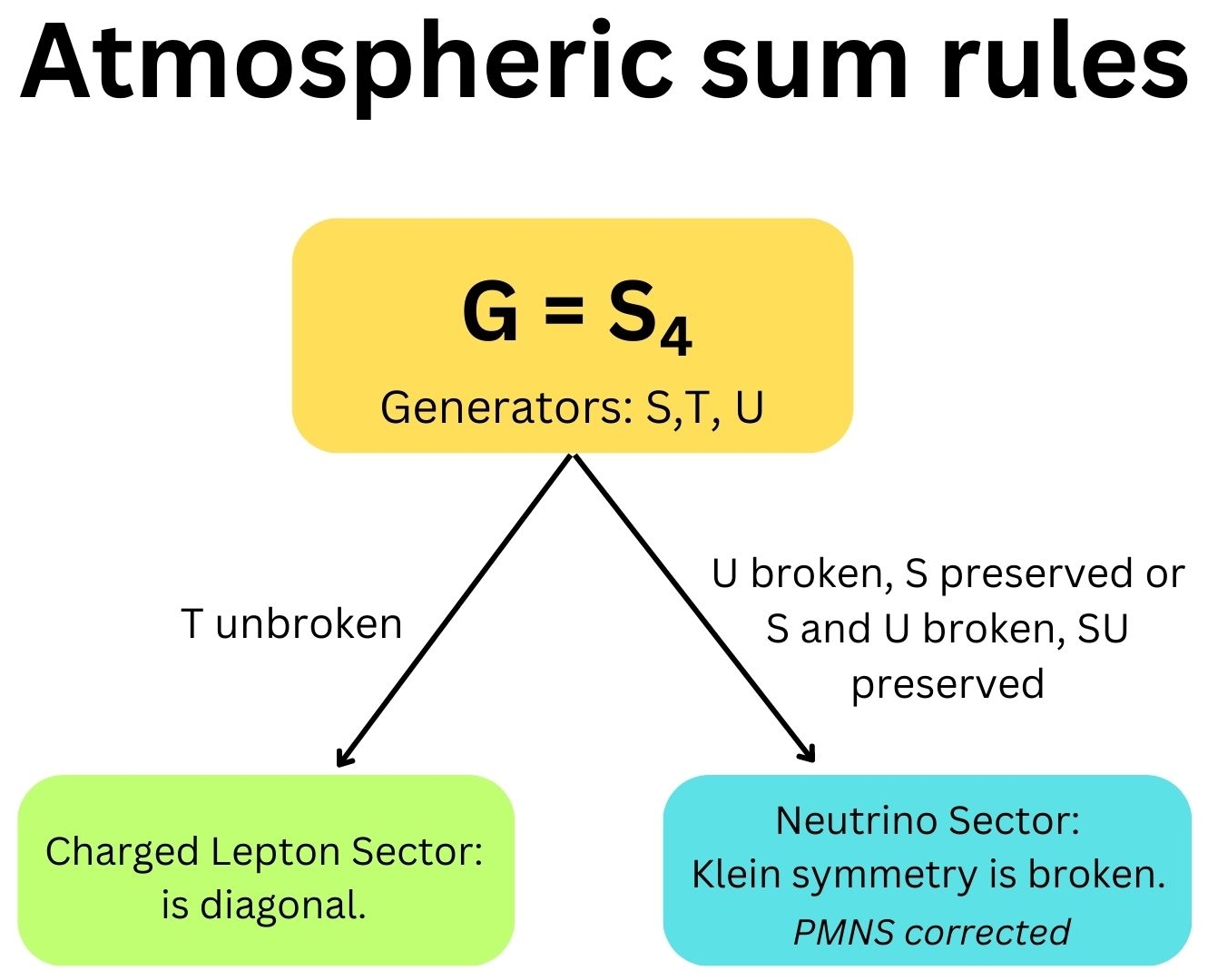}
		%
	\end{subfigure}
	\caption{\small In order to generate a non-zero (13) PMNS element, one or more of the generators $S,T,U$ must be broken. In the left panel we depict $T$ breaking leading to charged lepton mixing corrections and possible 
\textit{solar} sum rules. In the right panel, $U$ is broken, while either $S$ or $SU$ is preserved leading to neutrino
mixing corrections and \textit{atmospheric} sum rules. }
	\label{fig:S4corrections}
\end{figure}
In order to explain the experimental results $G$ needs to be broken and generate a non-zero $(13)$ PMNS element. This will lead to corrections to the leading order PMNS predictions from the discrete group $G$. In Figure \ref{fig:S4corrections} we illustrate two possible direction we can proceed to do that. The first one is to break the T generator while the Klein symmetry in the neutrino sector is exact (left hand side). This means that the charged lepton matrix is approximately diagonal. In the mass basis we will have then a correction to the neutrino mixing matrix by a unitary matrix $V_{e}$ and the PMNS is now $U_{\mathrm{PMNS}} = V_{e}^{\dagger} V_{\nu}$. Applying this to a group $G$ will lead to \textit{solar} sum rules.
The second direction is to preserve $Z_3^T$ but breaking $Z_2^U$ while keeping either $Z_2^{SU}$ or $Z_2^{S}$ unbroken (right hand side). This leads to corrections to the prediction of $G$ within the neutrino mixing and to \textit{atmospheric} sum rules.
It is convenient to introduce small parameters that can simplify the sum rules expressions and help us understand their physical behaviour since both in \textit{solar} and \textit{atmospheric} sum rules we implement a small deviation from the prediction of the exact finite discrete symmetries.
 We can consider the deviation parameters $s,r,a$ \cite{King:2007pr}

\begin{align}
	\sin \theta_{12} \equiv \frac{1+s}{\sqrt{3}}, \quad \sin \theta_{13} \equiv \frac{r}{\sqrt{2}}, \quad \sin \theta_{23} \equiv \frac{1+a}{\sqrt{2}},
	\label{eq:parameters}
\end{align}
that highlight the differences from TB mixing. Given the latest fit the $3\sigma$ allowed range for the solar, reactor and atmospheric deviation are respectively

\begin{align}
	\begin{gathered}
			-0.0999 <\; s < 0.0117, \\
		0.20146 <\; r < 0.21855, \\
		-0.0985 <\; a < 0.1129.
	\end{gathered}
\end{align}
This shows that the reactor angle differs from zero significantly ($r\neq 0$), but the solar and atmospheric angles remain consistent with TB mixing ($s=a=0$) at $3\sigma$. From a theoretical point of view, one of the goals of the neutrino experiments would be to exclude the TB prediction $s=a=0$ \cite{King:2012vj}, which is so far still allowed at $3\sigma$.


\section{Solar mixing sum rules}
\label{chap:3}

The first  possibility to generate a non-zero reactor angle, whilst maintaining some of the predictivity 
of the original mixing patterns,  
is to allow the the charged lepton sector to give a mixing correction to the leading order mixing matrix $U_{\nu}$.
This will lead to the so-called \textit{solar} sum rules, that are relations between the parameters that can be tested. This operation is equivalent to considering the $T$ generator of the $S_4$ symmetry which enforces the 
charged lepton mass matrix to be diagonal (in our basis) 
to be broken.

When the $T$ generator is broken, the charged lepton matrix is not exactly diagonal and it will give a correction to the PMNS matrix predicted by the symmetry group $G$. For example for the $S_4$, $U_{\mathrm{PMNS}}$ is not exactly $U_{\mathrm{TB}}$ but it receives a correction that we will compute. The fact that $S$ and $U$ are preserved leads to a set of correlations among the physical parameters, the \textit{solar} sum rules which are the prediction of the model.
For the \textit{solar} sum rules we can obtain a prediction for $\cos \delta$ as we shall now show. 

For example consider the case of TB neutrino mixing with  
the charged lepton mixing corrections involving only (1,2) mixing,
so that the PMNS matrix in Eq.~\eqref{Eq:PMNS_Definition} is given by,

\begin{equation}
\!
U_{\mathrm{PMNS}} = \left(\begin{array}{ccc}
\!c^e_{12}& s^e_{12}e^{-i\delta^e_{12}}&0\!\\
\!-s^e_{12}e^{i\delta^e_{12}}&c^e_{12} &0\!\\
\!0&0&1\!
\end{array}
\right)
\left( \begin{array}{ccc}
\sqrt{\frac{2}{3}} & \frac{1}{\sqrt{3}} & 0 \\
-\frac{1}{\sqrt{6}}  & \frac{1}{\sqrt{3}} & \frac{1}{\sqrt{2}} \\
\frac{1}{\sqrt{6}}  & -\frac{1}{\sqrt{3}} & \frac{1}{\sqrt{2}}
\end{array}
\right)
= \left(\begin{array}{ccc}
\! \cdots & \ \ 
\! \cdots&
\! \frac{s^e_{12}}{\sqrt{2}}e^{-i\delta^e_{12}} \\
\! \cdots
& \ \
\! \cdots
&
\! \frac{c^e_{12}}{\sqrt{2}}
\!\\
\frac{1}{\sqrt{6}}  & -\frac{1}{\sqrt{3}} & \frac{1}{\sqrt{2}}
\end{array}
\right)
\label{Ucorr}
\end{equation} 
The elements of the PMNS matrix are clearly related by \cite{Antusch:2007rk,Ballett:2014dua}

\begin{align}
	\frac{|U_{\tau1}|}{|U_{\tau2}|}=\frac{s_{12}^{\nu}}{c_{12}^{\nu}}=t_{12}^{\nu}= \frac{1}{\sqrt{2}}.
\label{solar}
\end{align} 
This relation is easy to understand if we consider only one charged lepton angle to be non-zero, $\theta^e_{12}$ then the third row of the PMS matrix in Eq.~\eqref{Ucorr} is unchanged, so the elements $U_{\tau i}$
may be identified with the corresponding elements in the uncorrected mixing matrix in Eq.\eqref{GR}.
Interestingly, the above relation still holds even if both $\theta^e_{12}$ and $\theta^e_{23}$ are non-zero.
However it fails if $\theta^e_{13}\ne 0$ \cite{Antusch:2022ufb}.

The above relation in Eq.~\eqref{solar} can be translated into a prediction for $\cos \delta$ as 
\cite{Ballett:2014dua}\footnote{See also \cite{Marzocca:2013cr}.}

\begin{align}
	\cos \delta=\frac{\tan \theta_{23} \sin \theta_{12}^2+\sin \theta_{13}^2 \cos \theta_{12}^2 / \tan \theta_{23}-(\sin \theta_{12}^{\nu })^2\left(\tan \theta_{23}+\sin \theta_{13}^2 / \tan \theta_{23}\right)}{\sin 2 \theta_{12} \sin \theta_{13}},
	\label{eq:solar-sum-rules}
\end{align}
where only the parameter $\sin \theta_{12}^{\nu}$ is model dependent and we have respectively $\sin \theta_{12}^{\nu} = 1/\sqrt{3}$, $\sin \theta_{12}^{\nu} =  1/\sqrt{2}$, $\tan \theta_{12}^{\nu} = 1/\varphi$ and $\theta_{12}^{\nu} = \pi/5$, $\cos \theta_{12}^{\nu} = \varphi/\sqrt{3}$ and $\theta_{12}^{\nu} = \pi/6$ for mixing based on TB, BM, GRa, GRb, GRc and HEX where $\varphi = (1+\sqrt{5})/2$.

\begin{figure}[t]
	\centering
	\begin{subfigure}{.5\textwidth}
		\centering
		\includegraphics[width=.9\linewidth]{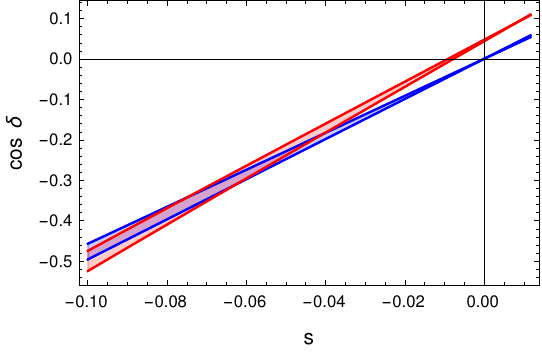}
		\label{fig:sub1}
	\end{subfigure}%
	\begin{subfigure}{.5\textwidth}
		\centering
		\includegraphics[width=.9\linewidth]{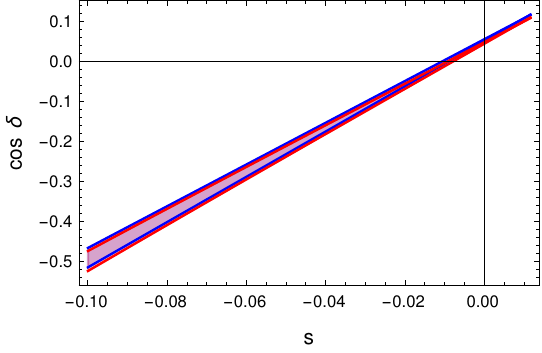}
		\label{fig:sub2}
	\end{subfigure}
	\caption{\small \textit{Solar} mixing sum rule predictions for TB neutrino mixing. In both panels the red band is the allowed region of the exact TB solar sum rules using the $3 \sigma$ range of r (i.e. the deviation of $\sin \theta_{13}$ from the TB value), it is plotted in the $3 \sigma$ range of s (i.e. the deviation of $\sin \theta_{12}$ from the TB value) and using the best fit value $a = 0.071$. The exact sum rules corresponds to Eq. \eqref{eq:solar-sum-rules}. Similarly blue band is the linearised sum rule allowed region which is given in Eq. \eqref{eq:solar-linear}. In the right panel the blue band is the second order expansion sum rule prediction, Eq. \eqref{eq:sum-rules-solar-second-order}, it matches the exact sum rule. }
	\label{fig:linearisation-solar}
\end{figure}

Let us discuss an approximation of the sum rules for the TB mixing as an example, where $\sin \theta_{12}^{\nu} = 1/\sqrt{3}$. We can re-write Eq. \eqref{eq:solar-sum-rules} using the parameters $s$, $a$ and $r$ defined in Eq. \eqref{eq:parameters} and then expand in them.  The linearised sum rule reads
\cite{King:2007pr}

\begin{align}
	\cos \delta = \frac{s}{r},
	\label{eq:solar-linear}
\end{align}
but it does not describe adequately the exact sum rules as shown in the left panel of Figure~\ref{fig:linearisation-solar}.
\begin{figure}[t]
	\centering
	\begin{subfigure}{.5\textwidth}
		\centering
		\includegraphics[width=.9\linewidth]{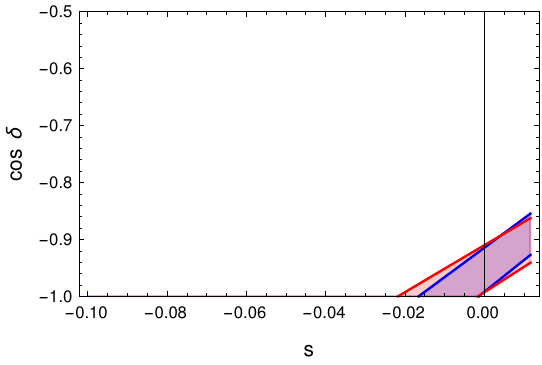}
		\label{fig:sub1}
	\end{subfigure}%
	\begin{subfigure}{.5\textwidth}
		\centering
		\includegraphics[width=.9\linewidth]{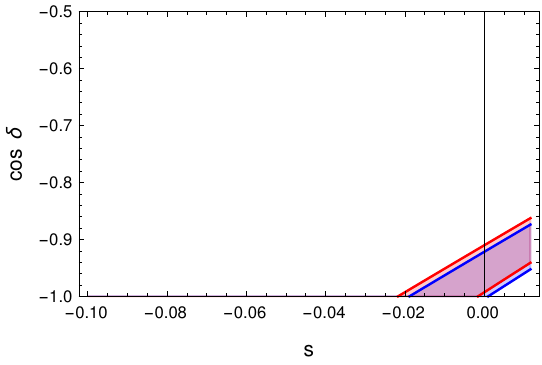}
		\label{fig:sub2}
	\end{subfigure}
	\caption{\small \textit{Solar} sum rule predictions for BM neutrino mixing. In the both panels the red band is the allowed region of the exact BM solar sum rules using the $3 \sigma$ range of r (i.e. the deviation of $\sin \theta_{13}$ from the TB value), it is plotted in the $3 \sigma$ range of s (i.e. the deviation of $\sin \theta_{12}$ from the TB value) and using the value $a = - 0.1$. The exact sum rules corresponds to Eq. \eqref{eq:solar-sum-rules}. Similarly blue band is the linearised sum rule allowed region which is given in Eq. \eqref{eq:solar-linear}. In the right panel the blue band is the second order expansion sum rule prediction, Eq. \eqref{eq:sum-rules-solar-second-order}, it matches the exact sum rule.	}
	\label{fig:linearisation-solar-2}
\end{figure}
Therefore we can go to the second order expansion, which is 

\begin{align}
	\cos \delta =\frac{s}{r}+\frac{r^2+8 a s }{4 r},
	\label{eq:sum-rules-solar-second-order}
\end{align}
and it matches the exact sum rule behaviour as seen on the right panel in Figure \ref{fig:linearisation-solar}. Similarly we can obtain higher order expansion for the other cases and check them against the data, like for the BM case showed in Figure \ref{fig:linearisation-solar-2}. In this case we did not choose the best fit value for $a$ because otherwise it would fall out of the physical range of $\cos \delta$ since BM is almost excluded by the data. The approximated expression for the sum rules can help us understand its behaviour and the dependence of $\cos \delta$ on the other parameters that are in general non-linear and assess the deviation from the non-corrected PMNS mixing. We then expect for the exact sum rules a first order linear dependence on $s$.
\begin{figure}[t]
	\centering
	\includegraphics[scale=0.70]{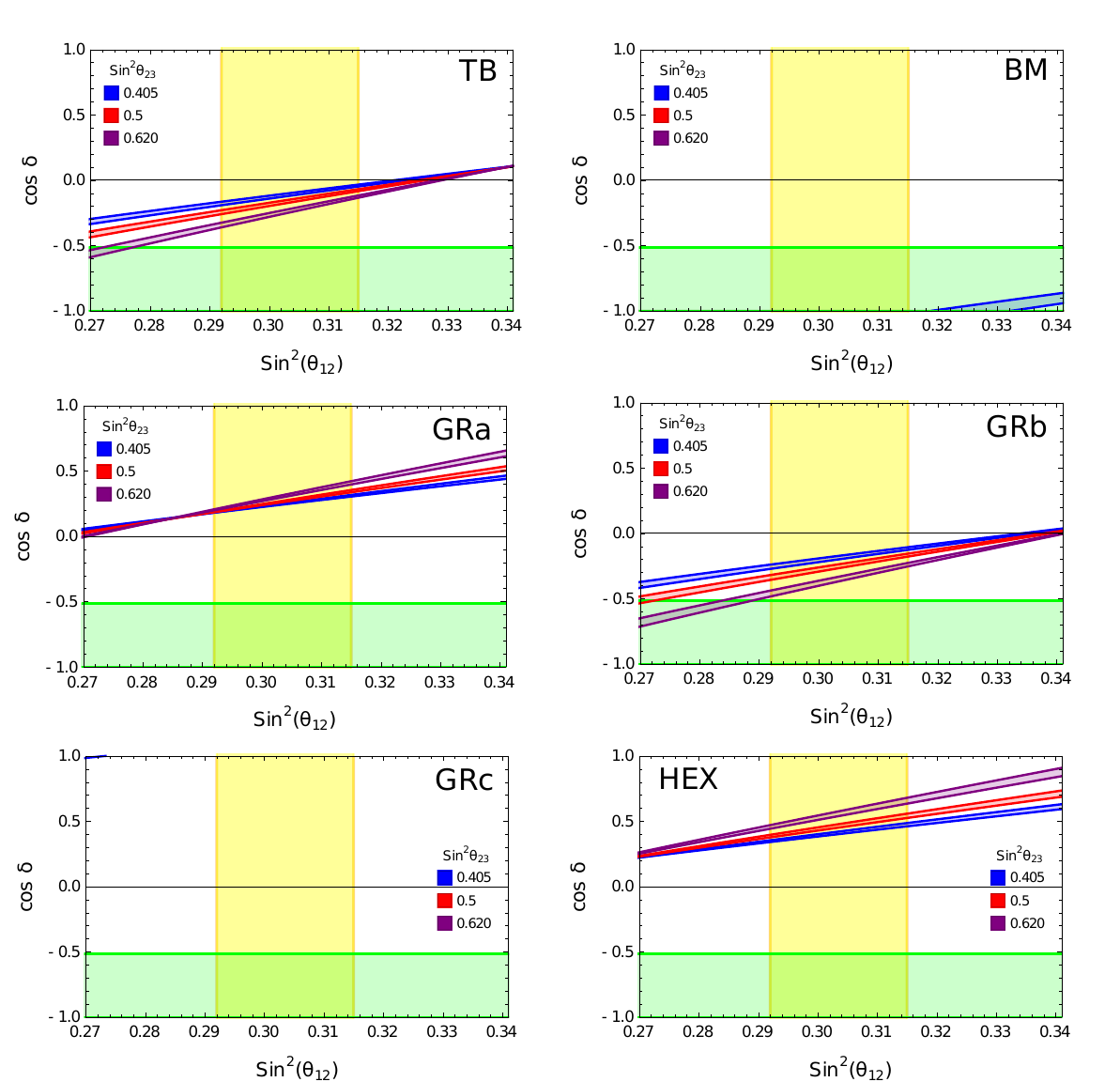}
	\caption{\small Summary of exact \textit{solar} sum rule predictions for different types of neutrino mixing.
	In the top left hand panel we present with the different colored band the sum rule prediction for TB for $\cos \delta$ letting $\sin \theta_{12}$ vary in its $3\sigma$ range, the different color denoted different choice of $\sin \theta_{23}$ given in the legend, in its $3\sigma$ range and the width of the band is given by the $3\sigma$ range in $\sin \theta_{13}$. The green and yellow band are the $1\sigma$ range for respectively $\cos \delta$ and $\sin \theta_{23}$. Similar plots for BM, GRa, GRb, GRc and HEX are presented respectively on the top right, center right, center left, bottom left, bottom right panels. The exact sum rules for the different models are derived from Eq. \eqref{eq:solar-sum-rules}. }
	\label{fig:SolarSumRules}
\end{figure}

In Figure \ref{fig:SolarSumRules} we present the exact sum rules prediction from Eq. \eqref{eq:solar-sum-rules} for TB, BM, GRa, GRb, GRc and HEX and the constraints from the fit of the neutrino oscillation data \cite{Esteban:2020cvm}. We require $\cos \delta$ to fall in the physical range $-1<\cos \delta<1$ and we present it in the y-axis. In all panel the x-axis is $\sin^2 \theta_{12}$ and the different colour bands are sampled in the allowed $\sin \theta_{23}$ region. The width of the band is given by allowing $\sin \theta_{13}$ to vary in its $3\sigma$ range. We notice that the $\theta_{12}^{\nu} = 45^{\circ}$ BM mixing (top-right panel) is closed to be excluded at $3\sigma$ and only low values of $\sin^2 \theta_{23}$ and high values of $\sin^2 \theta_{12}$ are still viable. Similarly for GRc mixing (bottom-left panel), with $\cos \theta_{12}^{\nu}=\varphi/3$, the viable parameter space is very tight, only for maximal values of $\sin \theta_{13}$ and minimal values of $\sin \theta_{12}$ and $\sin \theta_{23}$ we can obtain physical results for the CP phase. For TB mixing (top-left panel) with $\sin \theta_{12}^{\nu} = 1/\sqrt{3}$ in the neutrino sector with charged lepton correction lead consistent results in all parameters space, with the prediction for $\cos \delta$ that shows an approximately linear dependence on $\sin^2 \theta_{12}$ as understood by the leading order term in the sum rules in Eq. \eqref{eq:solar-linear}. 
The yellow and green bands are the $1\sigma$ range respectively of $\sin^2 \theta_{12}$ and $\cos  \delta$ and we notice how these ranges favor TB and GRb (with $\theta_{12}^{\nu} = \pi/5$ mixing) mixing. For both these models we see that the prediction of $\cos  \delta$ are mostly in the negative plane.
We finally notice that GRa (with $\tan \theta_{12}^{\nu} = 1/\varphi $) and HEX (with $\theta_{12}^{\nu} = \pi/6$) are the only models predicting mostly positive and physical values of $\cos  \delta$. Of the mixing pattern we studied GRb is favoured by the current $1\sigma$ ranges and BM and GRc are much disfavoured and only consistent with the far corners of the parameter space with a prediction of $|\cos  \delta\;| \approx 1$.


\section{Atmospheric mixing sum rules}
\label{chap:4}

In this section we discuss the second possibility, that is to have the T generator unbroken, therefore the charged lepton mixing matrix is exactly diagonal. In this case the correction to the PMNS matrix predicted from the group $G$ comes from the neutrino sector and it provides a non zero reactor angle. For each group there are two possible corrections achieved either breaking U and preserving S or with S and U broken and SU preserved. Therefore for each discrete symmetry we will study two mixing pattern \cite{Ballett:2013wya,Hernandez:2012ra,Luhn:2013lkn}.

\begin{figure}
	\centering
	\begin{subfigure}{.5\textwidth}
		\centering
		\includegraphics[width=.9\linewidth]{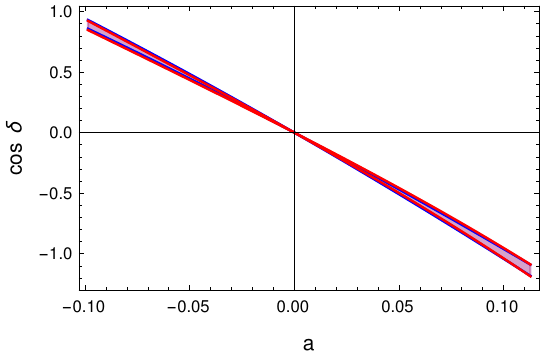}
		\label{fig:sub1}
	\end{subfigure}%
	\begin{subfigure}{.5\textwidth}
		\centering
		\includegraphics[width=.9\linewidth]{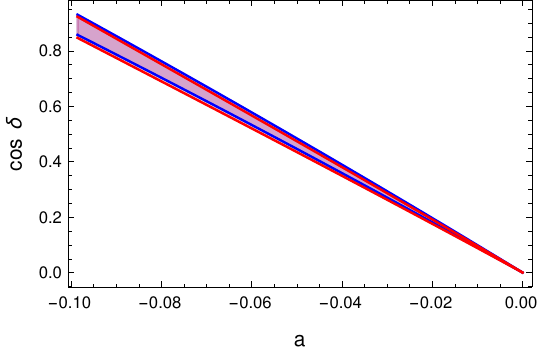}
		\label{fig:sub2}
	\end{subfigure}
	\caption{The red band is the allowed region of the exact TM2 sum rules using the 3$\sigma$ range of r and a (i.e. the deviation of $\sin \theta_{13}$ and $\sin \theta_{23}$ from the TB value), and it corresponds to Eq. \eqref{eq:atm-sum-rules}. The blue band is given by the second order sum rule which  is given in Eq. \eqref{eq:TM2-sumrules-second-order}. On the right we zoom on the region $-0.1<a<0$. }
	\label{fig:first-order-2}
\end{figure}

Let us consider again $G=S_4$ and the TB mixing in Eq. \eqref{TB}  as an example. 
If we break $S$ and $U$ but preserve $SU$ the first column of the TB matrix is preserved and we have the so-called TM1 mixing pattern \cite{Albright:2008rp,Albright:2010ap}
\begin{align}
	U_{\mathrm{TM} 1} \approx\left(\begin{array}{ccc}
		\sqrt{\frac{2}{3}} \; & -\; & - \\
		-\frac{1}{\sqrt{6}}\; & -\; & - \\
		\frac{1}{\sqrt{6}} \; & -\; & -
	\end{array}\right),
	\label{eq:TM1}
\end{align}
if instead $S$ is unbroken the second column is preserved and we have the second mixing pattern TM2 
\begin{align}
	U_{\mathrm{TM} 2} \approx\left(\begin{array}{ccc}
		- &\sqrt{\frac{1}{3}} & -  \\
		-& \sqrt{\frac{1}{3}} & -  \\
		- &- \sqrt{\frac{1}{3}}& - 
	\end{array}\right).
\end{align}
We can explicitly check this noticing that 

\begin{align}
		S \, \left(\begin{array}{c}
			\sqrt{\frac{1}{3}}  \\
			\sqrt{\frac{1}{3}}  \\
			 \sqrt{\frac{1}{3}}
		\end{array}\right) = \left(\begin{array}{c}
		\sqrt{\frac{1}{3}}  \\
		\sqrt{\frac{1}{3}}  \\
		\sqrt{\frac{1}{3}}
		\end{array}\right),
\end{align}
meaning that the second column of the TB mixing matrix is an eigenvector of the S matrix. Similarly for the first column with the $S U$ matrix.
%
%
In this second case where the second column of TB matrix is conserved we have

\begin{align}
	\left|U_{e 2}\right|=\left|U_{\mu 2}\right|=\left|U_{\tau 2}\right|=\frac{1}{\sqrt{3}},
\end{align}
and given the parametrisation in Equation \eqref{eq:parametrisation} we have 
\begin{align}
	\left|U_{e 2}\right|=|s_{12}& c_{13}|, \quad \left|U_{\mu 2}\right|=|c_{12} c_{23}-s_{12} s_{13} s_{23} e^{i \delta}|,\\  &\left|U_{\tau 2}\right|=|-c_{12} s_{23}-s_{12} s_{13} c_{23} e^{i \delta}|.
\end{align}
\begin{figure}[t]
	\centering
	\begin{subfigure}{.5\textwidth}
		\centering
		\includegraphics[width=.9\linewidth]{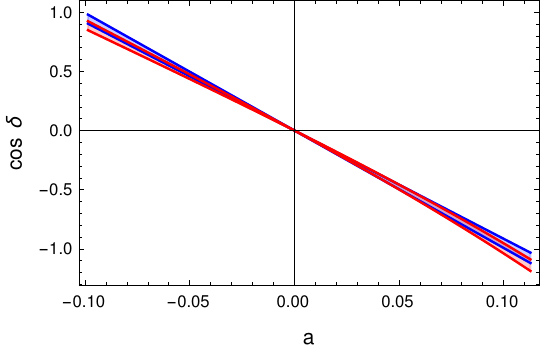}
		\label{fig:sub1}
	\end{subfigure}%
	\begin{subfigure}{.5\textwidth}
		\centering
		\includegraphics[width=.9\linewidth]{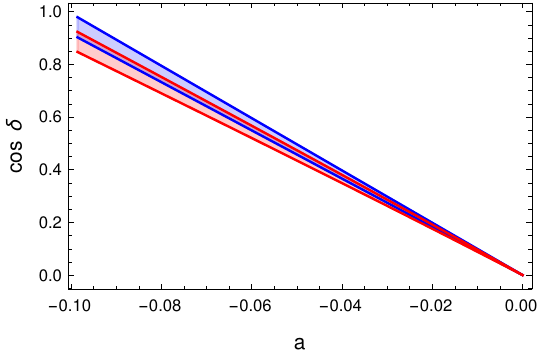}
		\label{fig:sub2}
	\end{subfigure}
	\caption{The red band is the allowed region of the exact TM2 sum rules using the 3$\sigma$ range of r and a (i.e. the deviation of $\sin \theta_{13}$ and $\sin \theta_{23}$ from the TB value), and it corresponds to Eq. \eqref{eq:atm-sum-rules}. The blue band is given by the linearised sum rule which  is given in Eq. \eqref{eq:TM2-sumrules-lin}. On the right we zoom on the region $-0.1<a<0$. }
	\label{fig:linearisation}
\end{figure}
Using the first equation $\left|U_{e 2}\right|=|s_{12} c_{13}|$ we have the first \textit{atmospheric} sum rule

\begin{align}	
	s_{12}^2 =  \frac{1}{3 \, c^2_{13}},
	\label{eq:atm-1}
\end{align}
that allows us to write $\theta_{12}$ in terms of $\theta_{13}$ and removing a parameter in our description and gives a prediction that can be tested. Using Eq. \eqref{eq:atm-1} and $|c_{12} c_{23}-s_{12} s_{13} s_{23} e^{i \delta}|^2=\frac{1}{3}$ we obtain the second \textit{atmospheric} sum rule \cite{Albright:2008rp,Albright:2010ap}

\begin{align}
	\cos \delta=\frac{2 c_{13} \cot 2 \theta_{23} \cot 2 \theta_{13}}{\sqrt{2-3 s_{13}^2}}.
	\label{eq:atm-sum-rules}
\end{align}
For the other models the discussion is similar where we call $X_1$ and $X_2$ the \textit{atmospheric} sum rules respectively derived by preserving the first and second column of the unbroken group with mixing $X$. In terms of the deviation parameters for TM2 we have the sum rule

\begin{align}
	\cos \delta= 	\frac{2 a(2+a) \left(-1+r^2\right)}{(1+a) \sqrt{1-2 a-a^2} r \sqrt{4-3 r^2}}.
	\label{eq:TM2-sum-rule}
\end{align}
We can expand this expression for small deviation parameters and at the zero-th order we have \cite{Ballett:2013wya}

\begin{align}
	\cos \delta = - \frac{2a}{r}
	\label{eq:TM2-sumrules-lin}
\end{align}
and in Figure \ref{fig:linearisation} we test this approximation against the exact sum rules using the experimental constraint in \eqref{eq:exp-data}. We can see that given the updated data the linear approximation is now insufficient to describe the exact expression as it was instead in previous studies \cite{Ballett:2013wya}. Similarly for TM1, as seen in Figure \ref{fig:first-order-2}. This is true for the other model we will discuss later and therefore we provide the higher order expansions that agrees with the exact sum rule in Eq. \eqref{eq:TM2-sum-rule} given the current data and is

\begin{align}
	\cos \delta = - \frac{2a}{r} - \frac{a^2}{r}
	\label{eq:TM2-sumrules-second-order}
\end{align}
For the TM2 example we see in Figure \ref{fig:linearisation} that the second order expansion is a good description of the exact sum rule. For TM1 instead, as shown in Figure \ref{fig:first-order-2} the third order expansion is needed.
\bgroup
\def\arraystretch{2}
\begin{table}[ht!]
	\centering
	\begin{tabular}{||c||c|c||}
		\hline \hline
		$  $     &  Exact sum rule &
		Approximated sum rule  \\ \hline
		TM1  & $\cos \delta=-\frac{\cot 2 \theta_{23}\left(1-5 \sin \theta_{13}^2\right)}{2 \sqrt{2} \sin \theta_{13} \sqrt{1-3 \sin \theta_{13}^2}}$ & $ \cos \delta = \frac{a}{r}+\frac{a^2}{2 r}+ \frac{2 a^3}{r}-\frac{7 a r}{4}$  \\ \hline
		TM2 & 	$\cos \delta=\frac{2 \cos \theta_{13} \cot 2 \theta_{23} \cot 2 \theta_{13}}{\sqrt{2-3 \sin \theta_{13}^2}} $ & $ \cos \delta =- \frac{2a}{r} - \frac{a^2}{r}$   \\ \hline
		GRa2 &  $\cos \delta=	 \frac{(1-\tan^{2} \theta_{23}) \csc \theta (1-3 \sin^2 \theta_{13} + (1 + \sin^2 \theta_{13}) \cos 2 \theta)}{8 \sin \theta_{13} \cos \theta_{23} \sqrt{\cos^2 \theta_{13} - \sin^2 \theta}}$ & $ \cos \delta =	a \frac{\sqrt{1+\cos 2\theta} \csc \theta}{r} \left( 1+\frac{a}{2} \right)$   \\ \hline \hline
	\end{tabular}
	\caption{Exact and approximated sum rules for the experimentally viable models, where $\theta = \arctan \frac{1}{\phi} $ and $\phi = \frac{1+\sqrt{5}}{2}$. 
	}
	\label{tab:sum-rules}
\end{table}
\egroup
%
\bgroup
\def\arraystretch{2}
\begin{table}[ht!]
	\centering
	\begin{tabular}{||c||c||c|c||}
		\hline \hline
		TM1 & $ \cos \theta_{12} = \sqrt{\frac{2}{3}} \frac{1}{\cos \theta_{13}}$ &   TM2 & $ \sin \theta_{12} = \frac{1}{\sqrt{3}\cos \theta_{13}}$  \\ \hline
		BM1  & $ \cos \theta_{12} = \frac{1}{\sqrt{2} \cos \theta_{13}} $ &	BM2  & $ \cos \theta_{12} = \frac{1}{\sqrt{2} \cos \theta_{13}} $ \\ \hline
		GRa1 & $ \cos \theta_{12} = \frac{\cos \theta}{\cos \theta_{13}}$ & GRa2 & $ \cos \theta_{12} = \frac{\sin \theta}{\cos \theta_{13}}$   \\ \hline
		GRb1 & $ \cos \theta_{12} = \frac{1+\sqrt{5}}{4\cos \theta_{13}}$   & GRb2 & $ \sin \theta_{12} = \frac{\sqrt{5+\sqrt{5}}}{4\cos \theta_{13}}$   \\ \hline
		GRc1 & $ \cos \theta_{12} = \frac{1+\sqrt{5}}{2 \sqrt{3}\cos \theta_{13}}$ &	GRc2 & $ \sin \theta_{12} = \frac{1+\sqrt{5}}{2\sqrt{3}\cos \theta_{13}}$   \\ \hline
		HEX1 & $ \cos \theta_{12} = \frac{\sqrt{3}}{2 \cos \theta_{13}}$  & HEX2 & $ \sin \theta_{12} = \frac{1}{2\sqrt{2}\cos \theta_{13}}$   \\ \hline
	\end{tabular}
	\caption{Exact sum rules plotted in Figure \ref{fig:SumRulesTM1-2}. 
where $\theta = \arctan \frac{1}{\phi} $ and $\phi = \frac{1+\sqrt{5}}{2}$.	}
	\label{tab:sum-rules2}
\end{table}
\egroup

\begin{figure}[]
	\includegraphics[width=0.85\linewidth]{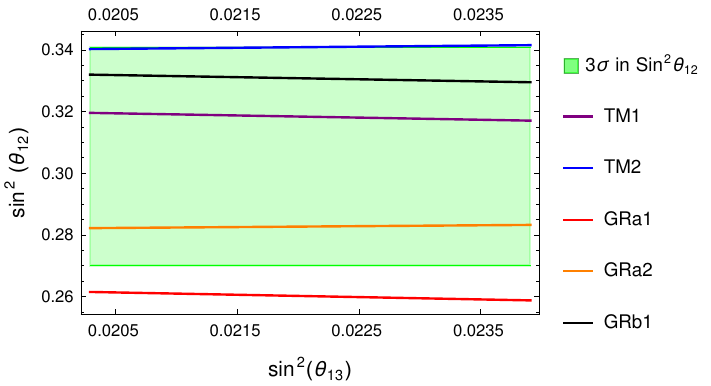}
	\caption{Summary of exact \textit{atmospheric} sum rule predictions which predict the solar angle
for different types of lepton mixing corresponding to a preserved column of the PMNS matrix, with only a mild dependence on the reactor angle. The corresponding Eqs. are collected in Table \ref{tab:sum-rules2}.
	The pink, blue, red, orange and black curves are respectively the predictions for TM1, TM2, GRa1, GRa2 and GRb1 mixing patterns. The $3\sigma$ allowed region is in green.}
	\label{fig:SumRulesTM1-2}
\end{figure}
\begin{figure}[]
	\includegraphics[width=0.85\linewidth]{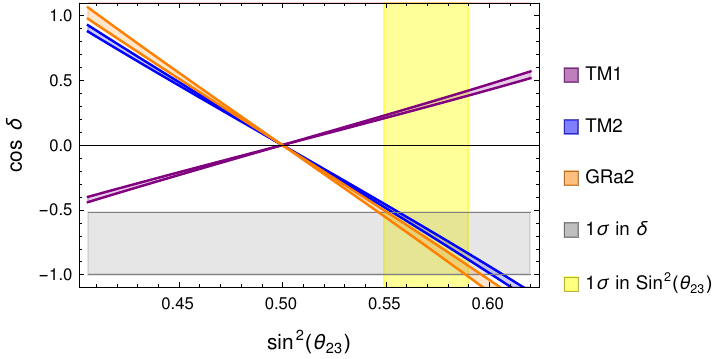}
	\caption{Summary of exact \textit{atmospheric} sum rule predictions which predict $\cos \delta$ in terms of the other mixing angles for different types of lepton mixing corresponding to a preserved column of the PMNS matrix. The corresponding Eqs. are collected in Table \ref{tab:sum-rules}.
	We present with the blue band the exact sum rule prediction for TM2 for $\cos \delta$ letting $\sin \theta_{13}$ vary in its $3\sigma$ range. In orange and purple we present the exact the sum rule predictions for GRa2 and TM1. The yellow and gray regions are respectively the $1\sigma$ range of $\sin \theta_{23}$ and $\cos \delta$, while the plot covers the whole $3\sigma$ range.}
	\label{fig:SumRulesTM1-1}
\end{figure}

Since the second exact sum rules are quite involved having an approximated expression is of help to understand the physical meaning of it and to understand the difference with respect to the TB model. We present in Table \ref{tab:sum-rules} the exact and approximated second sum rule for TM1, TM2 and GRa2 that as we will see later are the viable atmospheric mixing. Note that the approximated lead to simple results for TM1 and TM2 because the parameters $a$, $r$ and $s$ are built as deviation parameters from the TB mixing and beyond the first order expansion may not bring new insight for other mixing. We present in Table \ref{tab:sum-rules2} the first \textit{atmospheric} sum rules used in Figure \ref{fig:SumRulesTM1-2}. These results were derived using the normal ordered data without SK atmospheric results, the discussion regarding linearisation is the same including SK or considering the inverted ordering since $\sin \theta_{13}$ is very constrained and it does not change much in the different case considered.

In Figures \ref{fig:SumRulesTM1-2} and \ref{fig:SumRulesTM1-1} we study the exact \textit{atmospheric} sum rules for models obtained modifying TB, BM, GRa, GRb, GRc and HEX. In Figure \ref{fig:SumRulesTM1-2} we present first \textit{atmospheric} sum rule in Table \ref{tab:sum-rules2}, where the green band is the $3\sigma$ range for $\sin^2 \theta_{12}$. The models that do not appear are already excluded and far from the $3\sigma$ region. Therefore BM1, BM2, GRa1, GRb2, GRc1, GRc2, HEX1 and HEX2 are already excluded. In red we show GRa1 that is excluded a $3\sigma$ and in blue TM2, that is still not excluded only in a narrow parameter space, for high values of the solar and atmospheric angle. TM1 is showed in purple, GRa2 in orange and GRb1 in black. 

In Figure \ref{fig:SumRulesTM1-1} we show the exact \textit{atmospheric} sum rules (Table \ref{tab:sum-rules}) and the corresponding equations for other models that are still allowed from Figure \ref{fig:SumRulesTM1-2}. We plot $\cos \delta$ against $\sin \theta_{23}$ and letting $\sin \theta_{13}$ vary in its $3\sigma$ range, this gives the width of the different bands, in yellow and gray respectively are the $1\sigma$ band for $\sin^2 \theta_{23}$ and $\cos \delta$. The GRb1 mixing do not appear in the plot because it lays in unphysical values of $\cos \delta$. In purple, blue and orange we present TM1, TM2 and GRa2. We can see that given the $1\sigma$ bands, the GRa2 mixing is favoured when considering normal ordering and without the SK data, since TM2 is allowed only on a small portion of the parameter space as shown in Figure \ref{fig:SumRulesTM1-2}.


\section{Littlest Seesaw}
\label{chap:5}
There are many mechanism proposed to explain the smallness of the neutrino masses and that remain consistent with the data. For example the type I seesaw mechanism can address the problem through the introduction of heavy right-handed neutrinos. However in general it contains too many parameters to make any predictions for the neutrino mass and mixing. The constrained sequential dominance (CSD) model is a very predictive minimal seesaw model with two right-handed neutrinos and one texture zero ~\cite{King:2005bj,Antusch:2011ic,King:2013iva,King:2015dvf,King:2016yvg,Ballett:2016yod,King:2018fqh,King:2013xba,King:2013hoa,Bjorkeroth:2014vha}. As discussed in the Introduction the CSD($n$) scheme assumes that the two columns of the Dirac neutrino mass matrix are proportional to $(0,1, -1)$ and $(1, n, 2-n)$ or $(1, 2-n, n)$ respectively in the RHN diagonal basis (or equivalently $(0,1, 1)$ and $(1, n, n-2)$ or $(1, n-2, n)$) where the parameter $n$ was initially assumed to be a positive integer, but in general may be a real number. For example the CSD($3$) (also called Littlest Seesaw model)~\cite{King:2013iva,King:2015dvf,King:2016yvg,Ballett:2016yod,King:2018fqh} can give rise to phenomenologically viable predictions for lepton mixing parameters and the two neutrino mass squared differences $\Delta m^2_{21}$ and $\Delta m^2_{31}$, corresponding to special constrained cases of lepton mixing which preserve the first column of the TB mixing matrix, namely TM1 and hence satisfy \textit{atmospheric} mixing sum rules.

The Littlest Seesaw (LS) mechanism is one of the most economic neutrino mass generation mechanism that is still consistent with the experimental neutrino data \cite{King:2013iva,King:2015dvf,King:2016yvg}. We will show that after the choice of a specific $n$ value, all the neutrino observables are fixed by two free parameters. Different values of $n$ can be realised by different discrete symmetry groups. The LS introduces two new Majorana right-handed (RH) neutrinos $N_R^{atm}$ and $N_R^{sol}$ that will be mostly responsible for providing the atmospheric and solar neutrino mass respectively and the lightest SM neutrino is approximately massless; this is the idea of sequential dominance (SD) of RH neutrinos combined with the requirement for the $N_R^{atm}$ - $\nu_e$ interaction to be zero \cite{King:2002nf}. The Majorana neutrino mass matrix is given by the standard type I seesaw equation

\begin{align}
	M^\nu=-m^D M_R^{-1} m^{D^T},
	\label{eq:seesaw}
\end{align}
where the RH neutrino mass matrix $M_R$ is a $2\times2$ diagonal matrix

\begin{align}
	M_R=\left(\begin{array}{cc}
		M_{\mathrm{atm}} & 0 \\
		0 & M_{\mathrm{sol}}
	\end{array}\right), \quad M_R^{-1}=\left(\begin{array}{cc}
		M_{\mathrm{atm}}^{-1} & 0 \\
		0 & M_{\mathrm{sol}}^{-1}
	\end{array}\right),
	\label{eq:RH-masse}
\end{align}
where the convention for the heavy Majorana neutrino mass matrix corresponds to the
Lagrangian term $-\frac{1}{2}\ol{\nu^c_R} M_R \nu_{R}$ (which is equivalent to $-\frac{1}{2}\nu^T_R M_R \nu_{R}$)
and the convention for the light Majorana neutrino mass matrix corresponds to the
Lagrangian term $-\frac{1}{2}\ol{\nu_L} M^{\nu} \nu_{L}^c $ as in Eq. \eqref{mLLnu} which follows after performing the seesaw mechanism in Eq. \eqref{eq:seesaw}~\cite{King:2013eh}.
\footnote{Note that our convention for $M^{\nu}$ is the complex conjugate of the matrix used in the MPT package \cite{Antusch:2005gp} and in other studies in the literature \cite{Ding:2018fyz,deMedeirosVarzielas:2022fbw}. As will become apparent, in the LS case $M^{\nu}$ contains only one complex phase $\eta$, meaning that going from one to convention to the other $\eta$ changes sign: $\eta \rightarrow -\eta$. }

The Dirac mass matrix in Left-Right (LR) convention is a $3 \times 2$ matrix with arbitrary entries

\begin{align}
	m^D=\left(\begin{array}{cc}
		d & a \\
		e & b \\
		f & c
	\end{array}\right), \quad\left(m^D\right)^T=\left(\begin{array}{ccc}
		d & e & f \\
		a & b & c
	\end{array}\right),
\end{align}
where the entries are the coupling between the Majorana RH neutrinos and the SM neutrinos. The first column describe the interaction of the neutrinos in the flavour basis with the atmospheric RH neutrino and the second with the solar RH neutrino. The SD assumptions are that $d=0$, $d \ll e, f$, and 

\begin{align}
	\frac{(e, f)^2}{M_{\mathrm{atm}}} \gg \frac{(a, b, c)^2}{M_{\mathrm{sol}}},
\end{align}
these, together with the choice that of the almost massless neutrino to be the first mass eigenstate $m_1$, leads to $m_3 \gg m_2$ and therefore a normal mass hierarchy. This description can be further constrained choosing exactly $e=f$, $b=n a$ and $c=(n-2)a$ giving a simplified Dirac matrix 

\begin{align}
	m^D=\left(\begin{array}{cc}
		0 & a \\
		e & n a \\
		e & (n-2) a
	\end{array}\right),
	\label{eq:Dirac-mass}
\end{align}
that is called constrained dominance sequence (CSD) for the real number $n$ \cite{King:2005bj,Antusch:2011ic,King:2013iva}. It has been shown that the reactor angle is \cite{King:2015dvf}

\begin{align}
	\theta_{13} \sim(n-1) \frac{\sqrt{2}}{3} \frac{m_2}{m_3},
\end{align}
therefore this can provide non-zero and positive angle for $n>1$ and also excludes already models with $n \geq 5$ since they do not fit the experimental value. The choice $n\approx 3$ provides good fits to the data as we shall discuss.
Following the literature we will refer to CSD($n$) models with $n\approx 3$ as Littlest Seesaw (LS) models \cite{King:2015dvf}. 

The LS Lagrangian unifies in one triplet of flavour symmetry the three families of electroweak lepton doublets while the two extra right-handed neutrinos, $N_R^{\mathrm{atm}}$ and $N_R^{\mathrm{sol}}$ are singlets and reads \cite{King:2015dvf}
\begin{align}
	\mathcal{L}=-y_{\mathrm{atm}} \bar{L} \cdot \phi_{\mathrm{atm}} N_R^{\mathrm{atm}}-y_{\mathrm{sol}} \bar{L} \cdot \phi_{\mathrm{sol}} N_R^{\mathrm{sol}}-\frac{1}{2} M_{\mathrm{atm}} N_R^{\mathrm{atm}} N_R^{\mathrm{atm}}-\frac{1}{2} M_{\mathrm{sol}} N_R^{\mathrm{sol}} N_R^{\mathrm{sol}}+ h . c . \; ,
\end{align}
which can be enforced by a $Z_3$ symmetry and where $\phi_{\mathrm{atm}}$ and $\phi_{\mathrm{sol}}$ can be either Higgs-like triplets under the flavour symmetry or a combination of Higgses electroweak doublets and flavons depending on the specific choice of symmetry to use. In both cases the alignment should follow 

\begin{align}
	\phi_{\mathrm{atm}}^T \propto(0,1,1), \quad \phi_{\mathrm{sol}}^T \propto(1, n, n-2),
	\label{eq:normal}
\end{align}
or

\begin{align}
	\phi_{\mathrm{atm}}^T \propto(0,1,1), \quad \phi_{\mathrm{sol}}^T \propto(1, n-2, n).
	\label{eq:flipped}
\end{align}
We will refer to the first possibility in Eq. \eqref{eq:normal} as the normal case \cite{King:2013iva,King:2015dvf} and the second, in Eq. \eqref{eq:flipped} as the flipped case~\cite{King:2016yvg}. The predictions for $n$ in the flipped case are related to the normal one by  

\begin{align}
	\tan \theta_{23} \rightarrow \cot \theta_{23} \quad ( \theta_{23} \rightarrow \pi - \theta_{23}) \quad \quad \& \quad \quad \delta \rightarrow \delta + \pi,
	\label{eq:relation}
\end{align}
therefore we will discuss them together as one single $n$ case.  

There is an equivalent convention that can be found in the literature \cite{deMedeirosVarzielas:2022fbw}, where the alignment is chosen to be

\begin{align}
	\phi_{\mathrm{atm}}^T \propto(0,1,-1), \quad \phi_{\mathrm{sol}}^T \propto(1, n,2- n).
	\label{eq:convention2}
\end{align}
or 

\begin{align}
	\phi_{\mathrm{atm}}^T \propto(0,1,-1), \quad \phi_{\mathrm{sol}}^T \propto(1, 2-n, n).
	\label{eq:convention2}
\end{align}
that leads to the same results as the previous two cases respectively. In the neutrino mass matrix there will appear a $(-1)$ factor that is only a non-physical phase that can therefore be neglected. In particular the case $n = 1+\sqrt{6}$ that can be obtained with modular symmetry in \cite{deMedeirosVarzielas:2022fbw}\footnote{Notice that \cite{deMedeirosVarzielas:2022fbw} uses the MPT convention for $M^\nu$, which is related to our convention by a complex conjugation.} is still $n=1+\sqrt{6}$ in our convention using the Eq. \eqref{eq:normal}. Meaning that the case $n = 1-\sqrt{6}$ is just the flipped of $n = 1+\sqrt{6}$ and not a new LS model. 
We will follow the derivation in \cite{King:2015dvf} and using Eq. \eqref{eq:normal} derive the flipped result with Eq. \eqref{eq:relation}.
We will consider LS models corresponding to CSD($n$) models
with $n\approx 3$, in particular $n=2.5$, $3$ and $1+\sqrt{6}\approx 3.45$, together with their flipped cases.

For the normal cases of CSD($n$) the mass matrix in the diagonal charged lepton basis is given by

\begin{align}
	m^\nu=m_a\left(\begin{array}{ccc}
		0 & 0 & 0 \\
		0 & 1 & 1 \\
		0 & 1 & 1
	\end{array}\right)+m_b e^{i \eta}\left(\begin{array}{ccc}
		1 & n & n-2 \\
		n & n^2 & n(n-2) \\
		n-2 & n(n-2) & (n-2)^2
	\end{array}\right),
	\label{eq:neutrino-mass}
\end{align}
where we used Eqs. \eqref{eq:seesaw}, \eqref{eq:RH-masse} and \eqref{eq:Dirac-mass} 

 \begin{align}
	m_a=\frac{|e|^2}{M_{\mathrm{atm}}} \quad \quad \quad m_b=\frac{|a|^2}{M_{\mathrm{sol}}},
\end{align} 
and the only relevant phase is $\eta= \arg(a/e)$. 
At this point we notice that, in the diagonal charged lepton mass basis which we are using,
the PMNS mixing matrix is fully specified by the choice of $n$ and the parameters $m_b/m_a$ and $\eta$.
Indeed it is possible to derive exact analytic results for the masses and mixing angles  \cite{King:2015dvf}, and hence
obtain the LS prediction for the neutrino oscillation observables.

We first observe that 

\begin{align}
	m^\nu\left(\begin{array}{c}
		\sqrt{\frac{2}{3}} \\
		-\sqrt{\frac{1}{6}}\\
		\sqrt{\frac{1}{6}}
	\end{array}\right)=\left(\begin{array}{l}
		0 \\
		0 \\
		0
	\end{array}\right),
\end{align}
where the vector $(\sqrt{\frac{2}{3}},-\sqrt{\frac{1}{6}},\sqrt{\frac{1}{6}})^T$ is the first column of the TB matrix in Eq.~\eqref{TB} and is then an eigenvector of the neutrino mass matrix with eigenvalue $0$ and it corresponds to the massless neutrino eigenstate. This means that for a generic $n$ we get a TM1 mixing, Eq. \eqref{eq:TM1}, where the first column of the TB matrix is preserved and the other two can change. Therefore we can think of the LS as a
special case of the \textit{atmospheric} sum rules for the TB mixing. Since the \textit{atmospheric} sum rules were derived only using the fact that the first column of the TB matrix is preserved all LS implementations also follow the TM1 sum rules in Eq. \eqref{eq:TM1}. Once we have noticed this it is clear that $m_{\nu}$ can be block diagonalised using the TB matrix

\begin{align}
	m_{\text {block }}^\nu=U_{\mathrm{TB}}^T m^\nu U_{\mathrm{TB}}=\left(\begin{array}{ccc}
		0 & 0 & 0 \\
		0 & x & y \\
		0 & y & z
	\end{array}\right),
\end{align}
with

\begin{align}
	x=3 m_b e^{i \eta}, \quad y=\sqrt{6} m_b e^{i \eta}(n-1), \quad z=|z| e^{i \phi_z}=2\left[m_a+m_b e^{i \eta}(n-1)^2\right].
\end{align}
Finally we diagonalise $	m_{\text {block }}^\nu$ to obtain a matrix of the form $\operatorname{diag}\left(0, m_2, m_3\right)$

\begin{align}
	U_{\text {block }}^T m_{\text {block }}^\nu U_{\text {block }}=P_{3 \nu}^* R_{23 \nu}^T P_{2 \nu}^* m_{\text {block }}^\nu P_{2 \nu}^* R_{23 \nu} P_{3 \nu}^*=m_{\text {diag }}^\nu=\operatorname{diag}\left(0, m_2, m_3\right),
\end{align}
where the matrix including the phases are

\begin{align}
	\begin{gathered}
		P_{2 \nu}=\left(\begin{array}{ccc}
			1 & 0 & 0 \\
			0 & e^{i \phi_2^\nu} & 0 \\
			0 & 0 & e^{i \phi_3^\nu}
		\end{array}\right), \\
		P_{3 \nu}=\left(\begin{array}{ccc}
			e^{i \omega_1^\nu} & 0 & 0 \\
			0 & e^{i \omega_2^\nu} & 0 \\
			0 & 0 & e^{i \omega_3^\nu}
		\end{array}\right),
	\end{gathered}
\end{align}
and the angle we use to diagonalise is

\begin{align}
	R_{23 \nu}=\left(\begin{array}{ccc}
		1 & 0 & 0 \\
		0 & \cos \theta_{23}^\nu & \sin \theta_{23}^\nu \\
		0 & -\sin \theta_{23}^\nu & \cos \theta_{23}^\nu
	\end{array}\right) \equiv\left(\begin{array}{ccc}
		1 & 0 & 0 \\
		0 & c_{23}^\nu & s_{23}^\nu \\
		0 & -s_{23}^\nu & c_{23}^\nu
	\end{array}\right),
\end{align}
with the angle being fully specified by the free parameters $m_b/m_a$ and $\eta$, given by

\begin{align}
	t \equiv \tan 2 \theta_{23}^\nu=\frac{2|y|}{|z| \cos (A-B)-|x| \cos B},
	\label{eq:t-parameter}
\end{align}
where

\begin{align}
	\tan B=\tan \left(\phi_3^\nu-\phi_2^\nu\right)=\frac{|z| \sin A}{|x|+|z| \cos A},
\end{align}
and

\begin{align}
	A=\phi_z-\eta=\arg \left[m_a+m_b e^{i \eta}(n-1)^2\right]-\eta.
\end{align}
Recall that the PMNS matrix is the combination of the charged lepton and neutrino mixing matrices 

\begin{align}
	U_{\text{PMNS}}=U_{E_L} U_{\nu_L}^{\dagger},
		\label{eq:LS-PMNS}
\end{align}
where the neutrino mixing matrix, as we showed, is the product of the TB matrix and the $U_{\mathrm{block}}$ matrices 

\begin{align}
	U_{\nu_L}=U_{\mathrm{block}}^T U_{\mathrm{TB}}^T.
\end{align}
Now we can compare the PMNS matrix for the LS model with the standard parametrisation in Eq. \eqref{eq:parametrisation} to extract the mixing angles
\begin{align}
	\begin{aligned}
		\sin \theta_{13} & =\frac{1}{\sqrt{3}} s_{23}^\nu=\frac{1}{\sqrt{6}}\left(1-\sqrt{\frac{1}{1+t^2}}\right)^{1 / 2}, \\
		\tan \theta_{12} & =\frac{1}{\sqrt{2}} c_{23}^\nu=\frac{1}{\sqrt{2}}\left(1-3 \sin ^2 \theta_{13}\right)^{1 / 2}, \\
		\tan \theta_{23} & =\frac{\left|\frac{e^{i B}}{\sqrt{2}} c_{23}^\nu+\frac{1}{\sqrt{3}} s_{23}^\nu\right|}{\left|\frac{e^{i B}}{\sqrt{2}} c_{23}^\nu-\frac{1}{\sqrt{3}} s_{23}^\nu\right|}=\frac{\left|1+\epsilon_{23}^\nu\right|}{\left|1-\epsilon_{23}^\nu\right|},
	\end{aligned}
	\label{eq:angles}
\end{align}
with

\begin{align}
	\epsilon_{23}^\nu \equiv \sqrt{\frac{2}{3}} \tan \theta_{23}^\nu e^{-i B}=\sqrt{\frac{2}{3}} t^{-1}\left[\sqrt{1+t^2}-1\right] e^{-i B}.
\end{align}
The neutrino masses can be computed from $	m_{\text {block }}^\nu$ and they are

\begin{align}
	H_{\text {block }}^\nu=m_{\text {block }}^\nu m_{\text {block }}^{\nu \dagger}=\left(\begin{array}{ccc}
		0 & 0 & 0 \\
		0 & |x|^2+|y|^2 & |x||y|+|y| e^{i \eta} z^* \\
		0 & |x||y|+|y| e^{-i \eta} z & |y|^2+|z|^2
	\end{array}\right),
\end{align}
and after diagonalisation we can extract the eigenvalues as a function of the LS model parameters

\begin{align}
	\begin{aligned}
		m_2^2+m_3^2 & =T \equiv|x|^2+2|y|^2+|z|^2, \\
		m_2^2 m_3^2 & =D \equiv|x|^2|z|^2+|y|^4-2|x||y|^2|z| \cos A,
	\end{aligned}
\end{align}
and finally

\begin{align}
	\begin{aligned}
		m_3^2 & =\frac{1}{2} T+\frac{1}{2} \sqrt{T^2-4 D}, \\
		m_2^2 & =D / m_3^2, \\
		m_1^2 & =0,
	\end{aligned}
\end{align}
For the CP phase $\delta$ we have the cosine sum rule

\begin{align}
	\cos \delta=-\frac{\cot 2 \theta_{23}\left(1-5 \sin \theta_{13}^2\right)}{2 \sqrt{2} \sin \theta_{13} \sqrt{1-3 \sin \theta_{13}^2}},
\end{align}
that is the same as for the TM1 mixing in Table \ref{tab:sum-rules}. This can be understood since the LS is a subset of TM1 as we noticed before when we showed that the first column of the TB matrix is an eigenvector of the LS neutrinos mass matrix. Notice that for the flipped case $\cos \delta$ changes sign (because $\theta_{23} \rightarrow \pi - \theta_{23}$). Further information on the CP phase can be extracted from the Jarlskog invariant, which has been computed for the LS models \cite{King:2015dvf,King:2016yvg}:

\begin{align}
	J=s_{12} c_{12} s_{13} c_{13}^2 s_{23} c_{23} \sin \delta=\mp \frac{24 m_a^3 m_b^3(n-1) \sin \eta}{m_3^2 m_2^2 \Delta m_{32}^2},
\end{align}
where the negative sign corresponds to the normal case and the positive sign to the flipped. This leads to the sum rules for
$\sin \delta$ for the respective cases
\begin{align}
	\sin \delta= \mp \frac{24 m_a^3 m_b^3(n-1) \sin \eta}{m_3^2 m_2^2 \Delta m_{32}^2 s_{12} c_{12} s_{13} c_{13}^2 s_{23} c_{23}}.
\end{align}
Notice that in this case the model is more predictive than the discrete symmetries and it predicts both sine and cosine fixing unambiguously the CP phase $\delta$. Both $\sin \delta$ and $\cos \delta$ change sign going from the normal to the flipped cases meaning $\delta \rightarrow \pi +\delta$ as anticipated before. 

The above analytic results emphasise the high predictivity of these models which, for a given choice of $n$, 
successfully predict all the nine neutrino oscillation observables (3 angles, 3 masses, 3 phases) in terms of three input parameters 
namely the effective real masses $m_a, m_b$ and the phase $\eta$, which are sufficient to determine the neutrino mass matrix
in Eq. \eqref{eq:neutrino-mass}, where these parameters appear in the above analytic formulas.
However one neutrino mass is predicted to be zero ($m_1=0$), corresponding to a predicted normal hierarchy,
so one Majorana phase is irrelevant. For the remaining seven observables (3 angles, 2 masses, 2 phases) the overall neutrino mass scale may be factored out, and the Majorana phase is hard to measure, so that in practice we shall focus on the five observables,
namely the 3 angles $\theta_{13}, \theta_{12}, \theta_{23}$, the mass squared ratio 
$\Delta m_{21}^2/ \Delta m_{31}^2=m_2^2/m_3^2$ and the CP violating Dirac phase $\delta$, which are fixed by the two
input parameters, the phase $\eta$ and the ratio of the masses $r=m_b/m_a$,
In practice, we shall take the two most accurately determined observables, $\Delta m_{21}^2/ \Delta m_{31}^2$
and $\theta_{13}$ to fix the input parameters $\eta$ and $r=m_b/m_a$ within a narrow range, resulting in accurate predictions
for the remaining observables $ \theta_{12}, \theta_{23}$ and the Dirac phase $\delta$.
In addition we could add the input parameter $n$ as a free parameter, but this, together with the constrained form of mass matrices,
will eventually be determined by the flavour model.
In particular successful LS model structure corresponding to CSD($n$) can emerge from a theory of flavour as has been discussed in the literature for $n=3$~\cite{King:2016yvg},
$n=2.5$~\cite{Ding:2018fyz} and more recently 
$n=1+\sqrt{6}\approx 3.45$~\cite{Chen:2019oey,Ding:2019gof,Ding:2021zbg,deMedeirosVarzielas:2022fbw,deAnda:2023udh}.

\begin{figure}
	\includegraphics[scale=0.4]{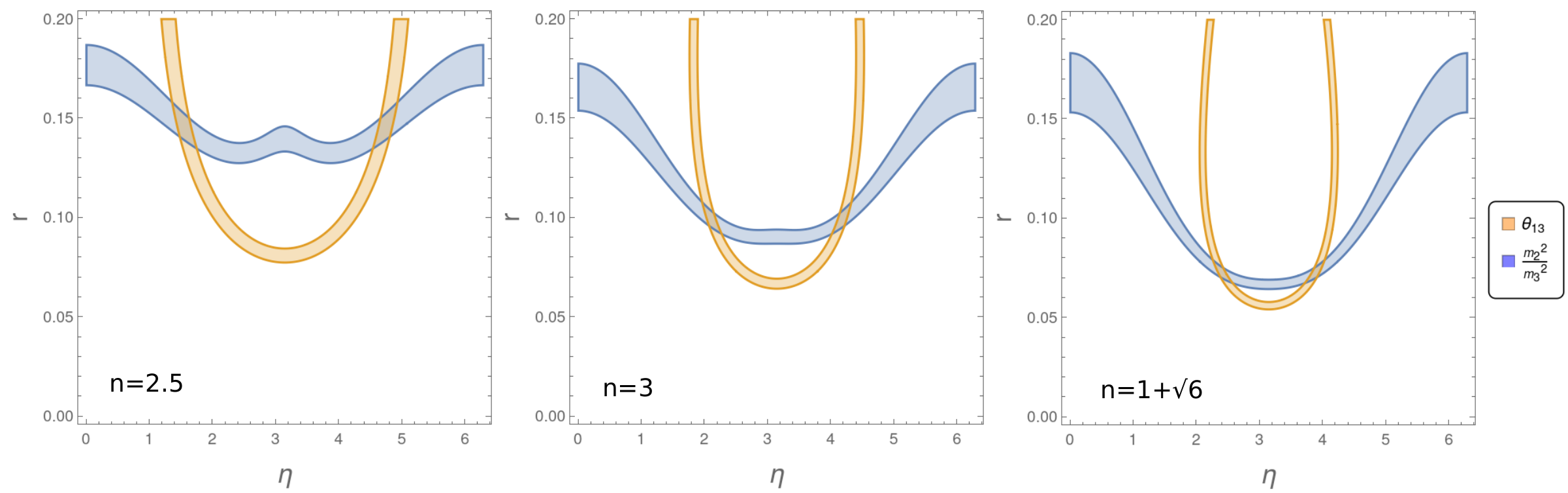}
	\caption{The results for the LS models with $n\approx 3$.
	The input parameters $\eta$ and $r=m_b/m_a$ are constrained to a good degree of accuracy by only two experimental observables, namely $\theta_{13}$ and the mass ratio $m_2^2/m_3^2$. The $3\sigma$ allowed region for $\theta_{13}$ and the mass ratio are respectively the blue and orange band. The area of intersection is the allowed parameter space for $\eta$ and $r$. From the left to the right we assume, $n=2.5$, $3$ and $1+\sqrt{6}\approx 3.45$.}
	\label{fig:LS}
\end{figure}

In Figure \ref{fig:LS} we consider the LS results for the above three cases with $n\approx 3$
and the corresponding flipped cases, which are all realised successfully via $S_4$ symmetry~\cite{King:2015dvf}. 
When we plot the experimental ranges of 
$\theta_{13}$ and the mass squared ratio $m_2^2/m_3^2$
in the $r - \eta$ plane, it is clear that only two small allowed parameter regions  
are allowed, which determine the maximal and minimal values of $r$ and $\eta$
as the intersection of the blue and orange bands.
Once we have the ranges of $r$ and $\eta$ for each value of $n$, thanks to the high predictivity of the model we can derive all the physical parameters and we can test them against the observed values. We do this for each value of
$n=3$, $1+\sqrt{6}\approx 3.45$ and $2.5$
in Tables \ref{tab:littlestn3} to \ref{tab:littlest2p5}. We do not present the plot for the flipped cases since they are exactly the same. In fact they involve only the mass ratio and $\theta_{13}$. %
\begin{table}[t]
	\centering
	\begin{tabular}{||c||c|c||c|c||}
		\hline \hline
		&  $n=3$   &  $\eta = 2.11 \pm 0.15$ &
		$\eta = 4.17 \pm 0.15$  & Exp. range \\ \hline
		& $\theta_{12}\; [^{\circ}]$  & $34.32{}^{+0.20}_{-0.24} $ & $34.32{}^{+0.20}_{-0.25}$  &   $31.31 - 35.74$ \\ \hline \hline
		normal & $\theta_{23}\; [^{\circ}]$  & $45.5{}^{+2.3}_{-2.4} $ & $45.5{}^{+2.3}_{-2.4}$  & $39.6 - 51.9$ \\ \hline
		normal & $\delta\; [^{\circ}]$ & $272.2^{+9.6}_{-11.0}  $ & $87.9 {}^{+11.0}_{-9.6} $  &  $0 -44 $ \& $108-360$  \\ \hline \hline
		flipped & $\theta_{23}\; [^{\circ}]$  & $44.5{}^{+2.3}_{-2.4} $ & $44.5{}^{+2.3}_{-2.4}$  & $39.6 - 51.9$ \\ \hline
		flipped & $\delta\; [^{\circ}]$ & $92.2^{+9.6}_{-11.0} $ & $267.9{}^{+11.0}_{-9.6}$  &  $0 -44 $ \& $108-360$  \\ \hline \hline
	\end{tabular}
	\caption{The LS predictions for $n=3$ where the two most accurately measured observables,
$\theta_{13}$ and the mass squared ratio $m_2^2/m_3^2$, are used to accurately determine the two input parameters 
$r=m_b/m_a= 0.100 \pm 0.008$ for two $\eta$ ranges as shown above, corresponding to the centre panel of 
Fig.~\ref{fig:LS}. This then leads to highly constrained predictions for the less accurately determined observables 
$\theta_{12}$, $\theta_{23}$ and $\delta$, which may be compared to the current experimental ranges as shown in the table.
All results are given to 3$\sigma$ accuracy.}
	\label{tab:littlestn3}
\end{table}
\begin{table}[t]
	\centering
	\begin{tabular}{|c||c||c|c||c||}
		\hline \hline
		& $  n = 1+ \sqrt{6}$     &  $\eta = 2.42 \pm 0.16$ &
		$\eta = 3.87 \pm 0.16$  & Exp. range \\ \hline
		& $\theta_{12}\; [^{\circ}]$  & $34.36^{+0.18}_{-0.21} $ & $34.36^{+0.18}_{-0.21} $ &  $31.31 - 35.74$  \\ \hline \hline
		normal & $\theta_{23}\; [^{\circ}]$  & $41.4^{+2.6}_{-2.6}$ & $41.5^{+2.6}_{-2.6}$  & $39.6 - 51.9$  \\ \hline
		normal& $\delta\; [^{\circ}]$ & $253.8{}^{+11.7}_{-13.8}$ & $105.7{}^{+13.7}_{-11.6}$  & $0 -44 $ \& $108-360$   \\ \hline \hline
		flipped & $\theta_{23}\; [^{\circ}]$  & $48.6^{+2.6}_{-2.6}$ & $48.5^{+2.6}_{-2.6}$  & $39.6 - 51.9$  \\ \hline
		flipped& $\delta\; [^{\circ}]$ & $74.8{}^{+11.7}_{-13.8}$ & $285.8^{+13.7}_{-11.6}$  & $0 -44 $ \& $108-360$   \\ \hline \hline
	\end{tabular}
	\caption{The LS predictions for $n = 1 + \sqrt{6} \approx 3.45$ 
	where the two most accurately measured observables,
$\theta_{13}$ and the mass squared ratio $m_2^2/m_3^2$, are used to accurately determine the two input parameters 
$r=m_b/m_a= 0.072 \pm 0.004$ for two $\eta$ ranges as shown above, corresponding to the right panel of 
Fig.~\ref{fig:LS}. This then leads to highly constrained predictions for the less accurately determined observables 
$\theta_{12}$, $\theta_{23}$ and $\delta$, which may be compared to the current experimental ranges as shown in the table.
All results are given to 3$\sigma$ accuracy.}
	\label{tab:littlestn1s6}
\end{table}
\begin{table}[t]
	\centering
	\begin{tabular}{||c||c||c|c||c||}
		\hline \hline
		 &$  n = 2.5$     &  $\eta = 1.5 \pm 0.2$ &
		$\eta = 4.7 \pm 0.2$  & Exp. range w/o SK  \\ \hline
		 & $\theta_{12}\; [^{\circ}]$  & $34.31^{+0.16}_{-0.20} $ & $34.28^{+0.17}_{-0.21}$  &   $31.31 - 35.74$ \\ \hline
		normal &$\theta_{23}\; [^{\circ}]$  & $51.5^{+1.9}_{-2.2} $ & $51.0^{+2.0}_{-2.3}$  & $39.6 - 51.9$  \\ \hline
	    normal &	$\delta\; [^{\circ}]$ & $299.9^{+9.2}_{-9.9}$ & $63.6^{+10.1}_{-9.3}$  &  $0 -44 $ \& $108-360$  \\ \hline \hline
		flipped &$\theta_{23}\; [^{\circ}]$  & $38.5^{+1.9}_{-2.2} $ & $39.0^{+2.0}_{-2.3}$  & $39.6 - 51.9$  \\ \hline
        flipped &	$\delta\; [^{\circ}]$ & $119.9^{+9.2}_{-9.9} $ & $243.6^{+10.1}_{-9.3}$  &  $0 -44 $ \& $108-360$  \\ \hline \hline
	\end{tabular}
	\caption{The LS predictions for $n = 2.5$ 
	where the two most accurately measured observables,
$\theta_{13}$ and the mass squared ratio $m_2^2/m_3^2$, are used to accurately determine the two input parameters 
$r=m_b/m_a=0.15 \pm 0.01$ for two $\eta$ ranges as shown above, corresponding to the left panel of 
Fig.~\ref{fig:LS}. This then leads to highly constrained predictions for the less accurately determined observables 
$\theta_{12}$, $\theta_{23}$ and $\delta$, which may be compared to the current experimental ranges as shown in the table.
All results are given to 3$\sigma$ accuracy.}
	\label{tab:littlest2p5}
\end{table}

In Table \ref{tab:littlestn3} we focus on the originally studied $n=3$ and its flipped case. We present the theoretical prediction and its uncertainty coming from the allowed region in Figure \ref{fig:LS} (centre panel) and the experimental bound. Since the theoretical prediction is exact given $\eta$ and $r$ we are allowing two significant figure for the theoretical errors. We notice that $\theta_{12}$ and $\theta_{23}$ fall well within the experimental range for all the cases and that even if $\delta$ is still not measured very precisely it allows us to exclude one of the two possible $\eta$ both in the normal and flipped case. In fact only the $\eta=2.11$ normal case and $\eta=4.17$ flipped case are within the $3\sigma$ experimental range. 

In Table \ref{tab:littlestn1s6} we focus on $n = 1 +\sqrt{6}\approx 3.45$, which can be realised with a modular symmetry \cite{deMedeirosVarzielas:2022fbw}, we notice that for the normal case both $\eta$ values are still allowed but with the $\delta$ prediction for $\eta=3.87$ that lie at the edge of the allowed experimental range. For the flipped case instead $\eta=2.42$ is excluded, thanks again to the bound on $\delta$. As before, in going from $n = 1 +\sqrt{6}$ to the flipped only changes the sign of $t$ in Eq. \eqref{eq:t-parameter}. The prediction for the mass ratio, $\theta_{13}$ and $\theta_{12}$ are independent of this sign while $\theta_{23}$ and $\delta$ are affected by it, as we can see in Eqs. \eqref{eq:angles} and as discussed above for 
$\delta$. The predictions are thus related by $\tan \theta_{23} \rightarrow \cot \theta_{23}$ (or $\theta_{23} \rightarrow \pi - \theta_{23}$) and $\delta \rightarrow \delta + \pi$. 

In Table \ref{tab:littlest2p5} we focus on $n=2.5$ and notice that, given the $\delta$ values, $\eta = 4.7$ is excluded for the normal case while for the flipped both $\eta$ values are allowed. Finally, $\theta_{23}$ lies in the higher and lower end of the experimental range respectively for the normal and flipped case making the $n=2.5$ disfavoured given the current data. This case is also known in the literature as $n=-1/2$ using the convention in Eq. \eqref{eq:convention2}. But it is more consistent to refer to it as $n=2.5$ in our notation. 

In summary, we see that most of the LS models with $n\approx 3$ are still allowed by current data. We have considered 
the cases $n=2.5$ and 
$n=1+\sqrt{6}\approx 3.45$ and compared the results to $n = 3$ which was the originally proposed CSD(3). We emphasise the high predictivity of the LS models which have three input parameters describing nine neutrino observables.
We have presented a new method here to present the results, namely to use the two most accurately measured observables,
$\theta_{13}$ and the mass squared ratio $\Delta m_2^2/ \Delta m_3^2=m_2^2/m_3^2$, to accurately constrain the two input parameters $r=m_b/m_a$
and $\eta$. This then leads to highly constrained predictions for the less accurately determined observables 
$\theta_{12}$, $\theta_{23}$ and $\delta$, which can be tested by future neutrino oscillation experiments.
Indeed already some of the possible LS cases are excluded by current data.
In addition all these LS cases predict zero lightest neutrino mass $m_1=0$, with a normal neutrino mass hierarchy,
and the neutrinoless double beta decay parameter $m_{\beta \beta}$ equal to $m_b$, which is just the first element of the neutrino mass
matrix in Eq. \eqref{eq:neutrino-mass}.
Indeed $m_{\beta \beta}=m_b$ can be readily determined from $\Delta m_2^2= m_2^2$, but its value is too small to be measured 
in the near future so we have not considered it here. On the other hand, a non-zero measurement of
$m_1$ or $m_{\beta \beta}$ in the inverted mass squared ordering region would immediately exclude the LS models.

\section{Conclusions}
\label{chap:6}

In the past decades many attempts have been made to explain the flavour structure of the PMNS matrix by imposing symmetry on the leptonic Lagrangian. These symmetries imply correlations among the parameters that are called sum rules. We have studied two types of sum rules: \textit{solar} and \textit{atmospheric} mixing sum rules. Then we have studied the littlest seesaw (LS) models
which obey the  TM1 \textit{atmospheric} mixing sum rule but are much more predictive. The goal of this paper has been to study all these approaches together in one place so that they may be compared, and to give an up to date analysis of the predictions of all of these possibilities, when confronted with the most recent global fits. 

In the case of \textit{solar} mixing sum rules, the $T$ generator of a given symmetry group is broken in the charged lepton sector in order to generate a non-zero reactor angle $\theta_{13}$. This leads with prediction for $\cos \delta$ that can be tested against the experimental data. These in turn show a preference for GRa and GRb mixing while BM and GRc are constrained to live in a very small window of the parameter space of current data. Future high precision neutrino oscillation experiments will constrain 
\textit{solar} mixing sum rules further as discussed elsewhere~\cite{Ballett:2014dua}. 

The \textit{atmospheric} mixing sum rules instead come from either the breaking of both $S$ and $U$ in the neutrino sector while preserving $SU$ or by breaking $S$ and preserving $U$. In this case we have two relations among the parameters that can be tested. We noticed that only TM1, TM2 and GRa2 are still allowed by the neutrino oscillation data with a preference for GRa2 and with TM2 very close to be excluded. Future high precision neutrino oscillation experiments will constrain 
\textit{atmospheric} mixing sum rules further as discussed elsewhere~\cite{Ballett:2013wya}.

We have also considered the class of 
LS models that follow the constrained sequential dominance idea, CSD($n$) with $n\approx 3$.
The LS models obey the TM1 \textit{atmospheric} mixing sum rule, 
but have other predictions as well.
We have compared the cases $n=2.5$, $n=3$ and $n=1+\sqrt{6}\approx 3.45$ which are predicted by theoretical models. 
These models are highly predictive with only two free real parameters fixing all the neutrino oscillation observables, making them candidates for being the most minimal predictive seesaw models of leptons still compatible with data. This is the first time that all three $n$ values above, both normal and flipped cases, 
have been studied together in one place, using the most up to date global fits.
We have also proposed a new way of analysing these models, which allows accurate predictions for the least well determined oscillation parameters $\theta_{12}$, $\theta_{23}$ and $\delta$ which we have shown to 
lie in relatively narrow 3$\sigma$ ranges, 
much smaller than current data ranges, but (largely) consistent with them, allowing these models to be decisively tested by future neutrino oscillation experiments, as has been discussed elsewhere~\cite{Ballett:2016yod}.
In our analysis we have ignored the model dependent renormalisation group (RG) corrections to LS models which have been shown to be generally quite small~\cite{Geib:2017bsw}.

In conclusion, we have shown that the recent global fits to experimental data have provided significantly improved constraints on all these symmetry based approaches, and future neutrino oscillation data will be able to significantly restrict the pool of viable models. 
In particular improvements in the measurement of the leptonic CP violating Dirac phase $\delta$ will strongly constrain all these cases. This is particularly true in LS models which provide very precise theoretical predictions for $\delta$, as well as  $\theta_{12}$ and $\theta_{23}$, consistent with current global fits.
Future precision neutrino experiments are of great importance to continue to narrow down the choice of possible PMNS flavour models based on symmetry
and lead to a deeper understanding of the flavour puzzle of the SM.

\acknowledgments

The work is supported 
by the European Union Horizon 2020 Research and Innovation programme under Marie Sklodowska-Curie grant agreement HIDDeN European ITN project (H2020- MSCA-ITN-2019//860881-HIDDeN).
SFK acknowledges the STFC Consolidated Grant ST/L000296/1.

\end{document}